\documentclass[pra,floatfix,twocolumn,showpacs]{revtex4}
\usepackage{amsmath,amssymb,amsfonts}
\usepackage{graphicx} % Include figure files
\usepackage{dcolumn} % Align table columns on decimal point
\usepackage{bm} % bold math

\newcommand{\erf}[1]{Eq.~(\ref{#1})}

\newcommand{\rt}[1]{\sqrt{#1}\,}

\newcommand{\op}[2]{\left|{#1}\rangle \langle{#2}\right|}

\newcommand{\an}[1]{\left\langle{#1}\right\rangle}
\newcommand{\ban}[1]{\big\langle{#1}\big\rangle}
\newcommand{\Ban}[1]{\Big\langle{#1}\Big\rangle}
\newcommand{\bTr}[1]{\Tr\big[{#1}\big]}

\newcommand{\tbt}[4]{\left(\begin{array}{cc} {#1} & {#2} \\[0.10cm] {#3} & {#4} \end{array}\right)}
\newcommand{\tbo}[2]{\left(\begin{array}{c} {#1} \\[0.05cm] {#2} \end{array}\right)}
\newcommand{\obt}[2]{\left(\begin{array}{cc} {#1} ,\; {#2} \end{array}\right)}

\newcommand{\vop}[1]{\hat{\bf {#1}}}
\newcommand{\vog}[1]{\hat{\bm{#1}}}

\newcommand{\smallsub}[1]{_{\text{\tiny #1}}}
\newcommand{\smallfrac}[2]{\mbox{$\frac{#1}{#2}$}}

\newcommand{\ID}[1]{{\rm I}_{\text{\tiny $#1$}}}
\newcommand{\mopID}[1]{\hat{\rm I}_{\text{\tiny $#1$}}}
\newcommand{\mbrac}[2]{\big\lfloor {#1},{#2} \big\rceil}
\newcommand{\sbrac}[2]{\big\lceil {#1}, {#2} \big\rfloor}

\newcommand{\beq}{\begin{equation}} 
\newcommand{\eeq}{\end{equation}}
\newcommand{\bqa}{\begin{eqnarray}} 
\newcommand{\eqa}{\end{eqnarray}}
\newcommand{\nn}{\nonumber}
\newcommand{\half}{\smallfrac{1}{2}}

\newcommand{\hp}{Hudson--Parthasarathy}
\newcommand{\ito}{It\^o} 

\newcommand{\sch}{Schr\"odinger} 
\newcommand{\hei}{Heisenberg}

\newcommand{\tp}{^{\top}} 
\newcommand{\dg}{^\dagger}
\newcommand{\ddg}{^\ddag}
\newcommand{\Tr}{{\rm Tr}}

\newcommand{\hx}{\hat{\bf x}}
\newcommand{\hc}{\hat{\bf c}} 

\newcommand{\rhoc}{\rho_{\rm c}}

\newcommand{\bin}{\vop{b}_{\rm in}}
\newcommand{\bout}{\vop{b}_{\rm out}}

\newcommand{\bmfb}{\vop{b}_{\rm outt}}
\newcommand{\muin}{\vog{\mu}_{\rm in}}
\newcommand{\mumfb}{\vog{\mu}_{\rm outt}}
\newcommand{\ym}{\vop{y}_1}
\newcommand{\ymfb}{\vop{y}_2}

\newcommand{\zetamfb}{\vog{\zeta}_{\rm outt}}

\newcommand{\Hext}{\hat{H}_{1}}
\newcommand{\Hm}{\hat{H}_{\rm m}}
\newcommand{\Hfb}{\hat{H}_{\rm fb}}
\newcommand{\Lm}{{\cal L}_{\rm m}}
\newcommand{\Lmfb}{{\cal L}_{\rm mfb}}

\newcommand{\Um}{\hat{U}_{\rm m}}
\newcommand{\Ufb}{\hat{U}_{\rm fb}}
\newcommand{\Umfb}{\hat{U}_{\rm mfb}}

\newcommand{\dBin}{d\vop{B}_{\rm in}}
\newcommand{\dBout}{d\vop{B}_{\rm out}}
\newcommand{\dBmt}{\dBout}
\newcommand{\dBmfb}{d\vop{B}_{\rm outt}}

\newcommand{\dUin}{d\vop{U}_{\rm in}}
\newcommand{\dUmfb}{d\vop{U}_{\rm outt}}

\newcommand{\dupsin}{d\vog{\upsilon}_{\rm in}}
\newcommand{\dupsmfb}{d\vog{\upsilon}_{\rm outt}}
\newcommand{\qctrl}{\hat{\bf u}}

\newcommand{\mrep}{{\sf M}}

\newcommand{\MdgM}{{\sf M}\dg {\sf M}}
\newcommand{\MMdg}{{\sf M} {\sf M}\dg}

\begin{document}

\title{The Quantum Theory of MIMO Markovian Feedback with Diffusive Measurements}

\author{A. Chia}
\author{H. M. Wiseman}
\affiliation{Centre for Quantum Computation and Communication Technology (Australian Research Council); \\
Centre for Quantum Dynamics, Griffith University, Brisbane, Queensland 4111, Australia.}

\date{\today}

\begin{abstract}

Feedback control engineers have been interested in MIMO (multiple-input multiple-output) extensions of SISO (single-input single-output) results of various kinds due to its rich mathematical structure and practical applications. An outstanding problem in quantum feedback control is the extension of the SISO theory of Markovian feedback by Wiseman and Milburn [Phys. Rev. Lett. {\bf 70}, 548 (1993)] to multiple inputs and multiple outputs. Here we generalize the SISO homodyne-mediated feedback theory to allow for multiple inputs, multiple outputs, and \emph{arbitrary} diffusive quantum measurements. We thus obtain a MIMO framework which resembles the SISO theory and whose additional mathematical structure is highlighted by the extensive use of vector-operator algebra. 

\end{abstract}


\maketitle

\section{Introduction}

Feedback control engineering \cite{Bec05} is ubiquitous in modern technology \cite{Ber02,FPE06}. As we further miniaturise technology, a quantum theory of feedback control can be expected to be essential \cite{HJM02,WM10}. In fact the realisation that quantum technology may benefit from modern control theory is currently driving a research program in which concepts from classical control systems \cite{DB08,ZDG96} are being applied and extended to quantum systems \cite{DDM+00,DHJ+00,DJ99,GJ09,JNP08,SP09,MP09,DP09,Mab09,YK03,JS08}. This facet of quantum feedback control makes it an interdisciplinary field, attracting both engineers and physicists.

\begin{figure}
\includegraphics[width=8.0cm]{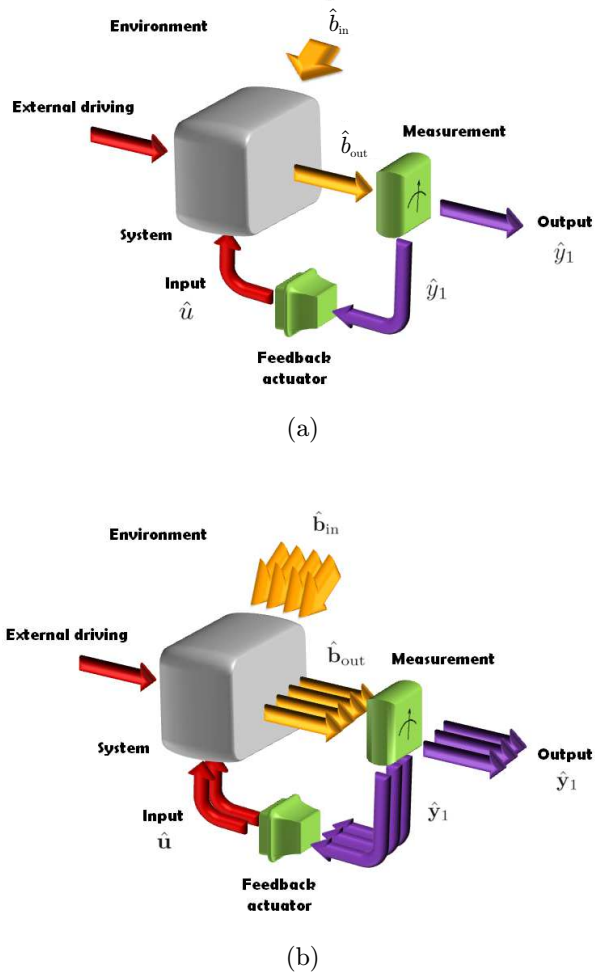} 
\caption{Markovian feedback in the case of (a) SISO, and (b) MIMO. Note the number of $\hat{b}_{\rm in}$ fields is not part of the definition of MIMO. For generality we will take there to be $L$ such inputs. We take the environment to be a collection of (bosonic) harmonic oscillators. The system interacts with the bath field $\bin$ and this process turns $\bin$ into $\bout$ which then gets detected. Here we are defining an $N$-component  vector-operator (vop) $\vop{a}$ as $\vop{a}\equiv(\hat{a}_1,\hat{a}_2,\ldots,\hat{a}_N)\tp$ (see Appendix A of Ref.~\cite{CW11a}). The detection process produces a current, modelled by $\ym$, which is then fed back into a feedback actuator. The actuator uses the information in the measured current to implement a control $\qctrl$ on the system. The measurement ouput $\ym$ is usually referred to as just the ``output'' and the control vop $\qctrl$ as the ``input''. It is possible to allow the number of outputs be different to the number of inputs, but for simplicity (and without loss of generality, see Appendix~\ref{EngineeringDefintion}) we will let these be the same, equal to $R$. Markovian feedback may then be defined by $\qctrl=\vop{y}_1$.}
\label{OverallScheme}
\end{figure}

A control strategy that has been widely studied is Markovian feedback \cite{WM10} which has useful applications in quantum information \cite{AWM03,TMW02,GTV96,HK97,CH07,WWM05}. This is a continuous (in time) process which can be briefly summarized by Fig.~\ref{OverallScheme}.
A general framework for such a process when the system has only one measurement output, one feedback input (Fig.~\ref{OverallScheme}(a), a case which we refer to as single-input single-output, abbreviated to SISO), mediated by homodyne detection was first put forth by Wiseman and Milburn \cite{WM93a}. In that work they treated feedback as an instantaneous process. A more detailed treatment that showed how to account for a feedback delay and how the limit of zero delay should be appropriately taken, giving rise to Markovian system evolution, was later given by Wiseman \cite{Wis94}. This is the most complete theory of Markovian feedback developed to date.

A theory of MIMO (multiple-input multiple-output, Fig.~\ref{OverallScheme}(b)) quantum feedback would be necessary in any situation where multiple degrees of freedom of a quantum system are monitored and controlled. The system could be a register of qubits, or the different canonical momenta (or positions) of a system of quantum objects. Indeed, investigations in this direction with a few inputs and outputs have already begun \cite{AWM03,ZWL+10,CWJ09,MW05,Man06,MW07}. With the drive to build  realistic quantum computing devices where quantum information would be encoded in many qubits a general theory of MIMO control would be an valuable tool to obtain.

The extension of Ref.~\cite{Wis94} to multiple inputs and multiple outputs would seem to be the obvious follow-up so it is natural to ask why this generalization was not made until now. There are two reasons for this. The first is related to the strategy underlying a master equation approach to open systems --- Changes in our distinguished system due to its interactions with other ancillary quantum systems are taken into account by including, in the master equation, parameters (numbers) which characterize these ancillary objects. The measurement step in the feedback loop shown in Fig.~\ref{OverallScheme} then defines a necessary point of interaction between the system and the measuring device. A mathematical representation of the measurement is therefore necessary; without it a master equation for the controlled system \emph{cannot} be derived. Finding this mathematical representation is nontrivial and it was not until 2001 that a representation of diffusive measurements with unit detection efficiency was found \cite{WD01}. The end result is a parameterization called the unravelling matrix, generalized in 2005 to include non-unit detection efficiency \cite{WD05}. In this paper we will use a different parameterization (which we have referred to as the M-rep \cite{CW11a}) because our results are simpler when expressed in terms of the M-representation of diffusive measurements.

The second reason for not extending the SISO work of Ref.~\cite{Wis94} to multiple inputs and multiple outputs earlier was due to a lack of motivation. The aforementioned research program of finding quantum-mechanical parallels of classical control has only proliferated in recent times \footnote{It is interesting to note that some engineers were already curious about such questions much earlier \cite{THC80,HTC83,OHT+84}.}. The physics and engineering communities at the time of Ref.~\cite{Wis94} were more or less separated and terms such as ``MIMO'' and ``nonlinear systems'' did not mean much to physicists. Control engineers have long been interested in generalizing various SISO results to the MIMO case, due to both its mathematical structure, and the prospects of practical applications that MIMO systems can offer \cite{Mac89,Gas08,SP05,Yin99,RG95,HSD99}. It remains to be an active line of research today in the engineering community \cite{TL03,LP03,Zha02,FSR04}. So a second motivation for constructing a MIMO theory of feedback is to allow quantum control to benefit from the works of engineers, and more generally, aid in the broader program of drawing analogies between classical and quantum theories of feedback control.

The paper is organized as follows: In Sec.~\ref{Background} we introduce the theory of quantum measurements in the \hei\ picture and discuss how such a model can be extended to include feedback. This theory is then used immediately in Sec.~\ref{UnconditionalDynamics} to describe the unconditional evolution of the system by deriving the Markovian MIMO feedback master equation and the Markovian quantum Langevin equation. There are two well-known approaches to obtaining these results and they are both discussed in Sec.~\ref{UnconditionalDynamics}. In Sec.~\ref{ConditionalEvolution} we consider time evolution with conditioning. The MIMO stochastic feedback master equation and two-time correlation function of the measured current are derived in this section. In Sec.~\ref{ConsistencyWithPrevResults} we show how our theory of MIMO feedback correctly reproduces previously known results in the limiting cases of homodyne- and heterodyne-mediated feedback. We then conclude with a discussion in Sec.~\ref{Discussion}.

At this point we would like to refer the reader to our exposition of vector-operator (or vop) algebra in the appendix of Ref.~\cite{CW11a} as this is used extensively in this paper. We also mention that for convenience we will not necessarily reflect the multi-component nature of vectors or vops in our language when they are referred to, such as in ``the field $\vop{a}$'', or, ``the current $\vop{y}$'', as opposed to using plurals as in ``the fields $\vop{a}$'' or ``the currents $\vop{y}$''.

%%%%%%%%%%%%%%%%%%%%%%%%%%%%%%%%%%%%%%%%%%%%%%%%%%%%%%%%%%%%%%%%%%%%%%%%%%%%%%%%%%%%%%%%%%%%%%%%%%%%%%%%%%%%%%%%%%%%%%%%%%%%%%%%%%%%%%%%%%%%%%%%%%%%%%%%

\section{Review of Heisenberg-Picture Dynamics}
\label{Background}

\subsection{Open Quantum Systems}

To set the premise of our theory we refer to Fig.~\ref{OverallScheme} but in the absence of the feedback actuator (i.e. $\vop{u}={\bf 0}$). The system and environment can be considered as one closed system whose time evolution is described by
\begin{align}
\label{HamiltonianWithoutFeedback}
	\hat{H} = \hat{H}_{0} + \hat{H}_{1} + \hat{H}_{\rm m} \;,
\end{align}
where $\hat{H}_{0}$ consists of the free Hamiltonians for the system and bath. Evolution due to external driving, or, for example, the extra Lamb shift that is often dropped in quantum optics \cite{Car02} are accounted for by $\hat{H}_{1}$. The environment is assumed to be a free bosonic field in one dimension (i.e. specified by a space-time coordinate) in the vacuum state and the system interacts with the environment by exchanging energy quanta with the bath field. We model this by the coupling Hamiltonian 
\begin{align}
\label{MmtHamiltonian}
	\hat{H}_{\rm m} = i ( \vop{b}\dg_{\rm in} \, \vop{c} - \vop{c}\dg \, \vop{b}_{\rm in} ) \;.
\end{align}
where $\hc$ and $\bin$ are each an $L$-component vop and the Hermitian conjugate of an $N$-vop $\vop{A}$ is defined by 
\begin{align}
	\vop{A}\dg = \big( \hat{A}\dg_1, \hat{A}\dg_2,\ldots,\hat{A}\dg_N \big)  \;.
\end{align}
Note that our measurement is performed on the bath, so within the standard quantum theory of indirect measurements \cite{BK92} the environment acts as our measuring apparatus and \eqref{MmtHamiltonian} effects a measurement interaction. The field $\vop{b}_{\rm in}(t)$ represents quantum noise and $d\vop{B}_{\rm in}(t) \equiv \vop{b}_{\rm in}(t) \, dt$ is a quantum Wiener increment \cite{Par92}. That is it has zero mean 
\begin{align}
\label{VacZeroMean}
	\big\langle d\vop{B}_{\rm in}(t) \big\rangle = {\bf 0} \;,
\end{align}
and satisfies the (quantum) \ito\ rule
\begin{align}
\label{QuantumItoRule}
	d\vop{B}_{\rm in}(t) \, d\vop{B}\dg_{\rm in}(t) = \hbar \, \mopID{L} \, dt \;,
\end{align}
with all other second or higher moments negligible. We are denoting an $L \times L$ identity mop (matrix-operator, see Ref.~\cite{CW11a}) by $\mopID{L}$. The dynamics due to $\hat{H}_0$ is usually well known and we can simplify matters by first transforming to a frame rotating at a frequency set by $\hat{H}_0$ and subsequently define all time evolution with respect to this frame. Unless required we will generally omit the time-dependence due to $\hat{H}_0$ and define our \sch\ and \hei\ pictures with respect to the rotating frame defined by $\hat{H}_0$ \cite{WM10}.

For simplicity we group $\Hext$ with $\Hm$ to define the time-evolution operator due to ``measurement'' by the \hp\ equation \cite{HP84}
\begin{align}
\label{MeasurementUnitary}
	\hbar \, d\Um(t,t_0) = {}& \big( -i \Hext \, dt - \half \, \hc\dg \hc \, dt  \nn \\
	                             & + \dBin\dg \;\! \hc - \hc\dg \dBin \big) \Um(t,t_0)  \;,
\end{align}
where $\Um(t_0,t_0)=\hat{1}$.

As a consequence of the singluar nature of $\bin$, the unitary evolution specified by \eqref{MeasurementUnitary} gives rise to an output field in the \hei\ picture 
\begin{align}
\label{MmtOutputField}
	\dBout(t) \equiv {}& \Um\dg(t+dt,t) \, \dBin(t) \, \Um(t+dt,t) \;, \nn \\
	               = {}& \hc(t) dt + \dBin(t) \;.
\end{align}
Note that $\bin$ and $\bout$ are different parts of the same quantum field, namely before and after interaction with the system \cite{GC85,CG84}. As such the input and output fields will only commute with an arbitrary system operator $\vop{s}$ at different times,
\begin{align}
\label{InputSysCommutator}
	\big\lfloor \dBin(t), \vop{s}(t') \big \rceil = 0 \quad \forall \; t' \le t \;, \\
\label{OutputSysCommutator}
	\big\lfloor \dBout(t), \vop{s}(t') \big\rceil = 0 \quad \forall \; t' > t  \;.
\end{align}
Here $\mbrac{\vop{A}}{\vop{B}}$ is the mop-bracket for two vops $\vop{A}$ and $\vop{B}$, defined as \cite{CW11a} 
\begin{align}
	\mbrac{\vop{A}}{\vop{B}} = \vop{A} \vop{B}\tp - \big( \vop{B} \vop{A}\tp \big)\tp  \;. 
\end{align}

An arbitrary vop $\vop{s}$ will evolve, due to the measurement interaction, according to the quantum Langevin equation derived from \eqref{MeasurementUnitary} 
\begin{align}
\label{QLEm}
	\hbar \, [ d\vop{s} \;\! ]_{\rm m} 
	= {}& \big( \, i \;\! [\hat{H}_{1},\vop{s}\;\!] +  {\cal J}[\hc\ddg] \;\! \vop{s} - \half \;\! \{ \hc\dg \hc, \vop{s} \} \, \big) dt \nn \\
	    & + \; [\;\!\hc\dg d\vop{B}_{\rm in} - d\vop{B}\dg_{\rm in} \hc, \vop{s}\;\!] \;,
\end{align}
where $\vop{A}\ddg \equiv \big( \vop{A}\tp \big)\dg$ and (see \cite{CW11a})
\begin{align}
	{\cal J}[\vop{A}\ddg] \, \vop{B} = \big( \vop{A}\ddg \vop{B}\tp \big)\tp \vop{A}  \;.
\end{align}
It is then easy to show that transforming this to the \sch\ picture gives the master equation due to measurement
\begin{align}
\label{MasterEqnWithoutFeedback}
	\hbar \, [d\rho\;\!]_{\rm m} \equiv {\cal L}_{\rm m} \;\! \rho \, dt = -i \;\! [\hat{H}_{1},\rho] dt + {\cal D}[\hc] \rho \, dt   \;,
\end{align}
where
\begin{align}
	{\cal D}[\hc] \rho = \hc\tp \rho \;\!\hc\ddg - \half \big\{ \hc\dg \hc, \rho \big\} \;,
\end{align}
and $\big\{\hat{A},\hat{B}\big\} \equiv \hat{A} \hat{B} + \hat{B} \hat{A}\,$.

\subsection{Quantum Measurements}
\label{DiffusiveOutput}

The output field $\bout$ is then measured and the detector produces a current ${\bf y}$. In the \hei\ picture the current is represented by a vector-operator, which in general will be some function of the output field $\bout$
\begin{align}
	\vop{y}_1 = g\big(\bout,\vog{\xi} \;\! \big) \;,
\end{align}
where $\vog{\xi}$ is measurement noise.

For the remainder of this paper we concentrate on the class of diffusive measurements. It was shown previously that the output of such a measurement can be represented by an $R \times 1$ vop \cite{CW11a,WM10} 
\begin{align}
\label{CurrentOperator}
	\hbar \, \vop{y}_1 \, dt = {\sf M}\dg \dBout + {\sf M}\tp \dBout\ddg  + \hbar \, d\vog{\upsilon}_{\rm in} \;.
\end{align}
The subscript for the current here does not mean that it is related to $\hat{H}_0$ and $\Hext$ in \eqref{HamiltonianWithoutFeedback}, instead it is to remind us that the current is defined in terms of output field $\dBout$. This will be useful when we consider feedback in Sec.~\ref{ZeroFeedbackDelayApproach} when the current will be defined in terms of the input field. Note that corresponding to each component of $\hc$ (or each dissipative channel) we need at most two quadrature measurements so $R \le 2L$. The matrix ${\sf M}$ is $L \times R$ defined by  
\begin{align}
\label{DefnOfM}
	\MMdg / \hbar  \in \mathfrak{H} \;,
\end{align}
where ${\mathfrak H} = \left\{ {\rm diag}({\bm \eta}\;\!) \, | \forall \, k, \, \eta_k \in [0,1] \;\! \right\}\;\!$. The noise $d\vog{\upsilon}_{\rm in}$ in \eqref{CurrentOperator} is a $R \times 1$ Hermitian vop with zero mean and correlations given by
\begin{align}
\label{DefnOfZ}
	 (\hbar \, d\vog{\upsilon}_{\rm in}) \, (\hbar \, d\vog{\upsilon}_{\rm in})\tp = \hbar \, {\sf Z} \, dt  \;,
\end{align}
where 
\begin{align}
\label{ZInTermsOfM}
	{\sf Z} = \hbar \, \ID{R} - \MdgM \;.
\end{align}
We can express $d\vog{\upsilon}_{\rm in}$ in terms of independent quantum Wiener increments 
\begin{align}
\label{dUAppear1}
	\hbar \, d\vog{\upsilon}_{\rm in} = \rt{\sf Z} \;\! \dUin + \rt{\sf Z}^{*} \;\! \dUin\ddg  \;.
\end{align}
The increments $\dUin$ are completely uncorrelated with the system so they satisfy 
\begin{align}
\label{dUAppear2}
	\mbrac{\dUin(t)}{\vop{s}(t')} = 0  \quad  \forall \; t, t'  \;.
\end{align}
We remind the reader that this is not what is usually referred to as the measurement noise $d\vop{v}_{\rm m}$.

\subsection{Adding Feedback}

We can describe feedback on the system by adding another Hamiltonian $\hat{H}_{\rm fb}$ to \eqref{HamiltonianWithoutFeedback}:
\begin{align}
\label{HamiltonianWithFeedback}
	\hat{H} = \hat{H}_0 + \hat{H}_1 + \hat{H}_{\rm m} + \hat{H}_{\rm fb} \;.
\end{align}
In general $\hat{H}_{\rm fb}$ will describe the coupling of the input $\qctrl$, which may be a functional of the current $\vop{y}_1$, to the system. Markovian feedback can be defined as the coupling of the measured current $\vop{y}_1$ (in which case $\qctrl=\vop{y}_1$) to a Hermitian system vop $\vop{f}$ (see Appendix~\ref{EngineeringDefintion}). As a result of working in the idealized limit where $\vop{y}_1$ contains white noise it is only sensible to consider $\vop{f}$ being coupled linearly to $\vop{y}_1$, i.e.
\begin{align}
\label{requiredfbHam}
	\hat{H}_{\rm fb} = \hbar \: \hat{\bf f}\tp \, \vop{y}_1 \;,
\end{align}
where $\vop{f}$ and $\vop{y}_1$ are at the same time. Coupling $\vop{f}$ to any nonlinear function of $\vop{y}_1$ would generate time evolution which is indescribable by (quantum) stochastic calculus.
%
%\begin{figure}
%\includegraphics[width=7cm]{FeedbackInteraction.eps} 
%\caption{Interactions defining Markovian feedback. From \eqref{CurrentOperator} coupling $\vop{f}$ to $\ym$ is equivalent to coupling $\vop{f}$ to the time-delayed fields $\rt{\sf Z} \muin(t-\tau)$ and ${\sf M}\bmt(t-\tau)$, where $\muin dt=\dUin$. Note that Markovian system dynamics emerges only when $\tau \to 0^{+}$.}
%\label{FeedbackInteraction}
%\end{figure}
%

The careful reader will notice a number of issues with the Hamiltonian (\ref{requiredfbHam}). First, $\vop{y}_1$ does not commute with $\vop{f}$ at the same time, so $\hat{H}_{\rm fb}$ as it stands is not even Hermitian. Second, it does not strictly exist because although $\vop{y}_1\,dt$ exists as a stochastic increment, $\vop{y}_1$ does not.

The first problem can be solved in two ways as was recognized in Ref.~\cite{Wis94}. The first is to realize that in actuality there must be a finite time delay in the feedback loop. Thus, strictly we have
\beq 
\label{requiredfbHamtau}
	\hat H_{\rm fb} =  {\hbar} \: \hat{\bf f}\tp \, \vop{y}_1(t-\tau) \,,
\eeq
and $\vop{y}_1(t-\tau)$ commutes with all system operators at times later than $t-\tau$ and so acts as a complex number for $\tau \ne 0$. The limit $\tau \to 0^+$ can be taken at the end of all calculations. We will derive a Markovian ($\tau \to 0^+$) master equation with feedback using this method in Sec.~\ref{MasterEquationNonzeroDelay}. The second approach is to treat the feedback as an instantaneous process at the outset by ensuring that the measurement acts before the feedback. We follow this approach in Sec.~\ref{ZeroFeedbackDelayApproach}.

The second issue is more serious, and for general (not necessarily linear) quantum systems care must be taken in determining the evolution generated by \erf{requiredfbHamtau}.

Our definition of Markovian feedback is directly in terms of the feedback Hamiltonian. Placing the definition on the Hamiltonian is sensible and appeals to physicists since the Hamiltonian is the generator of time evolution. In Appendix~\ref{EngineeringDefintion} we define feedback in a manner that draws upon the traditional control systems approach. In this language one can differentiate between system dynamics that is linear and nonlinear and the results of this paper can be seen to apply to the more general (nonlinear) regime.

%\begin{figure}
%\includegraphics[width=8.4cm]{OverallSchemeSchPic.eps} 
%\caption{Schematic of some photoemissive system which couples to the environment via some set of operators $\hc$. The time evolution of the system based on knowledge of the noisy signal ${\bf y}$ also follows a noisy path but in Hilbert space which can be described by a stochastic master equation. If one ignores the measured result ${\bf y}$, or equivalently do not measure, then the system state evolves according to the master equation. This corresponds to averaging over the stochastic evolution described in the former case by a classical probability distribution for the measurement noise.}
%\label{OverallSchemeSchPic}
%\end{figure}

                        %%%%%%%%%%%%%%%%%%%%%%%%%%%    Markovian Master Equation for Diffusive Unravellings    %%%%%%%%%%%%%%%%%%%%%%%%%

\section{Unconditional Dynamics}
\label{UnconditionalDynamics}

\subsection{Diffusion-mediated Feedback Starting with Non-zero Feedback Delay}
\label{MasterEquationNonzeroDelay}

\subsubsection{Feedback master equation}

Since we have already introduced the most general form of a master equation in the absence of feedback \eqref{MasterEqnWithoutFeedback}, we will only derive ${\cal L}_{\rm fb}$ in 
\begin{align}
	\hbar \, \dot{\rho} = \big( {\cal L}_{\rm m} + {\cal L}_{\rm fb} \big) \rho \;.
\end{align}
We will start in the Heisenberg picture in which case the quantum Langevin equation corresponding to $\dot{\rho}$ is 
\begin{align}
\label{dsDefinition}
	d \hat{\bf s} = [d\hat{\bf s}\;\!]_{\rm m} + [d\hat{\bf s}\;\!]_{\rm fb} \;,
\end{align}
where $[d\hat{\bf s}]_{\rm m}$ is given by \erf{QLEm}. The feedback contribution $[d\hat{\bf s}\;\!]_{\rm fb}$ can be obtained from 
\begin{align}
\label{dsfb}
	[d\hat{\bf s}\;\!]_{\rm fb} = \hat{U}\dg_{\rm fb}(t+dt,t) \, \hat{\bf s} \, \hat{U}_{\rm fb}(t+dt,t) - \hat{\bf s} \,. 
\end{align}
The unitary operator here is given by
\begin{align}
\label{UfbDefn}
	\hat{U}_{\rm fb}(t+dt,t) = {}& e^{-i \hat{H}_{\rm fb} dt / \hbar} \;.
\end{align}
It is perhaps not entirely obvious that deriving ${\cal L}_{\rm fb}$ (from either the \sch\ or \hei\ picture) and adding it to ${\cal L}_{\rm m}$ should result in the correct master equation since $[d\vop{s}]_{\rm m}$ and $[d\vop{s}]_{\rm fb}$ are defined with different time-evolution operators. This is justified in Appendix~\ref{LangevinEqnExplained}. Expanding \eqref{UfbDefn} to order $dt$,
\begin{align}
\label{Ufb}                        
	\hat{U}_{\rm fb}(t+dt,t) = \hat{1} - i \, \hat{\bf f}\tp \vop{y}_1(t-\tau) dt - \smallfrac{1}{2} \,\hat{\bf f}\tp \hat{\bf f} \, dt \;. 
\end{align}
We have used the \ito\ rule to obtain the last term in \erf{Ufb}. Substituting \erf{Ufb} into \eqref{dsfb}, retaining only terms of order $dt$, and multiplying by $\hbar$ we obtain
\begin{align}
\label{QLEfb}
	\hbar \, [d\vop{s}\;\!]_{\rm fb} = {}& - i \hbar \, \vop{s} \, [ \;\! \hat{\bf f}\tp \vop{y}_1(t-\tau) dt \,] 
                   	                     + i \hbar \, [ \;\! \vop{f}\tp \vop{y}_1(t-\tau) dt \, ] \, \vop{s}  \nn \\
	                                     & + \, \hbar \, {\cal D}[\vop{f}] \;\! \vop{s} \, dt \,. 
\end{align}
The bath is assumed to be in the vacuum state so the initial joint system-bath state is 
\begin{align}
\label{SysBathState}
	\rho\smallsub{SB} = \rho \otimes \op{\bf 0}{\bf 0} \;,
\end{align}
where $\rho$ is the system state and $\op{\bf 0}{\bf 0}$ the bath state. Remember that we are in the \hei\ picture so $\rho\smallsub{SB}$ does not evolve. To derive a master equation for $\rho$ we will take the ensemble average of \eqref{QLEfb} with respect to $\rho\smallsub{SB}$ and this immediately eliminates the vacuum noise contained in $\vop{y}_1$ since the vacuum inputs are completely independent of the system. This also suggests that we should normally order the terms containing $\dBout$ (since then $\dBin$ will annihilate the vacuum to give zero when averaged). Considering the first term of \eqref{QLEfb} for the moment, we obtain, upon substituting in \eqref{CurrentOperator} 
\begin{align}
\label{FirstTermOfQLE}
	- i \hbar \, \big\langle \vop{s} \, [ \,\hat{\bf f}\tp \vop{y}_1(t-\tau) \;\! dt \, ] \,\big\rangle 
	=  {}& - i \;\! \big\langle \;\! \vop{s} \, [\;\!\vop{f}\tp {\sf M}\dg \dBout(t-\tau)]  \nn \\
       & + \vop{s} \, [\;\!\vop{f}\tp {\sf M}\tp \dBout\ddg(t-\tau)] \;\! \big\rangle  \,. 
\end{align}
The first term here is already in normal order while the second term can be written as
\begin{align}
\label{Reordering}
	\vop{s} \, [\;\!\vop{f}\tp {\sf M}\tp \dBout\ddg(t-\tau)] = {}& [\vop{s} \, \dBout\dg(t-\tau)] \, {\sf M} \,\vop{f}\tp  \nn \\
                                                            = {}& [\dBout\ddg(t-\tau) \, \vop{s}\tp]\tp {\sf M} \, \vop{f} \;,
\end{align}
where we have noted the mop-bracket \eqref{OutputSysCommutator} in \eqref{Reordering}. Using these orderings and \eqref{MmtOutputField}, the average of \eqref{FirstTermOfQLE} is simply
\begin{align}
	- i \hbar \, \big\langle \vop{s} \, [ \,\hat{\bf f}\tp \hat{\bf y}(t-\tau) \;\! dt \, ] \,\big\rangle 
	=  {}& - i \;\! \big\langle \;\! \vop{s} \, [\;\!\vop{f}\tp {\sf M}\dg \hc(t-\tau)]  \nn \\
       & + [\;\!\hc\ddg(t-\tau) \, \vop{s}\tp]\tp {\sf M} \, \vop{f} \;\! \big\rangle  \,. 	
\end{align}
Now taking the Markovian ($\tau \to 0$) limit and writing the average as a trace we get
\begin{align}
	& \hspace{-1cm} -i\hbar \, \big\langle \vop{s} \, [ \,\vop{f}\tp \vop{y}_1 \;\! dt \, ] \, \big\rangle  \nn \\
	= {}& -i \, {\rm Tr} \Big\{ \vop{s} \, \vop{f}\tp {\sf M}\dg \hc \;\! \rho\smallsub{SB} 
	                            + \big(\hc\ddg \;\! \vop{s}\tp\big)^{\!\top} {\sf M} \, \vop{f} \;\! \rho\smallsub{SB} \Big\} \, dt  \nn \\
	= {}& -i \, {\rm Tr} \Big\{ \vop{s} \, \vop{f}\tp {\sf M}\dg \hc \;\! \rho\smallsub{SB} 
	                            + \vop{s} \;\! \big( {\sf M} \, \vop{f} \;\! \rho\smallsub{SB} \big)^{\!\!\top} \hc\ddg  \Big\} \, dt  \nn \\ 
\label{FeedbackOne}
	= {}& -i \, {\rm Tr} \Big\{ \vop{s} \, \vop{f}\tp \big( {\sf M}\dg \hc \;\! \rho\smallsub{SB} 
	                            + \rho\smallsub{SB} \;\! {\sf M}\tp \;\! \hc\ddg \big) \Big\} \, dt \;.	                            
\end{align}

To obtain the Markovian limit of the average of the second term in \eqref{QLEfb} we can perform a similar calculation as above, or, alternatively note that
\begin{align}
	& \hspace{-0.5cm} \lim_{\tau \to 0^+} i \hbar \, \big\langle \, [\;\! \vop{y}_1\tp\!(t-\tau) \;\! dt \: \vop{f} \;\!] \, \vop{s} \, \big\rangle  \nn \\
	{}& =  \Big\{ \lim_{\tau \to 0^+} - i \hbar \, \big\langle \, \vop{s}\dg \, [\;\! \vop{f}\tp \vop{y}_1(t-\tau) \, dt \;\!] \, \big\rangle \Big\}\dg              \nn \\
\label{FeedbackTwo}
	{}& = i \, {\rm Tr} \Big\{ \vop{s} \;\! \big( \rho\smallsub{SB} \, \hc\dg \;\! {\sf M} \, \vop{f} 
	                           +  \hc\tp \, {\sf M}^* \rho\smallsub{SB} \big) \vop{f}\Big\} \, dt  \;.   
\end{align}
The last line is obtained by letting $\vop{s} \to \vop{s}\dg$ in \eqref{FeedbackOne} and using the cyclic property of trace to permute $\vop{s}$ to the left. The average of ${\cal D}[\vop{f}]\vop{s}\;\!dt$ in \eqref{QLEfb} can simply be expressed as 
\begin{align}
\label{FeedbackThree}
	\hbar \, \big\langle {\cal D}[\vop{f}] \;\! \vop{s} \big\rangle \, dt 
	= {\rm Tr} \Big\{ \vop{s} \: \hbar \, {\cal D}[\vop{f}] \;\! \rho\smallsub{SB} \Big\} \, dt  \;.
\end{align}
Adding \eqref{FeedbackOne}, \eqref{FeedbackTwo}, and \eqref{FeedbackThree}, we arrive at  
\begin{align}
\label{FeedbackQLE}
	\hbar \, \big\langle \;\! [d\vop{s}]_{\rm fb} \big \rangle 
	= {}& {\rm Tr} \Big\{ \vop{s} \Big[ \hbar \, {\cal D}[\vop{f}] \, \rho\smallsub{SB} 
	      - i \, \vop{f}\tp \big( {\sf M}\dg \hc \;\! \rho\smallsub{SB} 
	      + \rho\smallsub{SB} \;\! {\sf M}\tp \;\! \hc\ddg \big)  \nn \\
	    & + i \big( \rho\smallsub{SB} \, \hc\;\!\dg \, {\sf M} 
	      + \hc\tp \, {\sf M}^* \, \rho\smallsub{SB} \big) \;\! \vop{f} \, \Big] \Big\} \, dt  \nn \\[0.2cm]
	= {}& {\rm Tr} \Big\{ \vop{s} \Big( \!-\!i \big\lceil \;\! \vop{f}, {\sf M}\dg \hc \;\! \rho\smallsub{SB} 
	      + \rho\smallsub{SB} \;\! {\sf M}\tp \hc\ddg \big\rfloor  \nn\\
	    & + \hbar \, {\cal D}[\vop{f}] \, \rho\smallsub{SB} \Big) \Big\} \, dt  \;.
\end{align}
In the last equality we have made use of the sop-bracket, defined by \cite{CW11a}
\begin{align}
	\sbrac{\vop{A}}{\vop{B}} = \vop{A}\tp \vop{B} - \vop{B}\tp \vop{A} \;.
\end{align}
Remember that we are only working out the time evolution due to feedback so the feedback contribution to the full master equation is defined by
\begin{align}
	\hbar \, \langle \;\! [d\vop{s}(t)]_{\rm fb} \rangle = {\rm Tr}\smallsub{S} \Big\{ \vop{s}(0) \, \hbar \, [d\rho(t)]_{\rm fb} \Big\}
\end{align}
where $\rho(t)$ here is defined by the partial trace over the bath $\rho(t) = {\rm Tr}\smallsub{B}\{\rho\smallsub{SB}(t)\}$. We thus obtain, in the \sch\ picture, where operators are understood to be time-independent and $\rho$ time-dependent,
\begin{align}
	  \hbar \, [d\rho\;\!]_{\rm fb} \equiv {}& {\cal L}_{\rm fb} \, \rho \, dt  \nn \\
\label{Lfb}
	  = {}& \Big( \hbar \;\! {\cal D}[\hat{\bf f}]\rho \,
	         - i \big\lceil \;\! \vop{f}, {\sf M}\dg \hc \;\! \rho + \rho \;\! {\sf M}\tp \hc\ddg \big\rfloor \Big) dt  \;.
\end{align}
Adding this to the measurement master equation defined by \eqref{MasterEqnWithoutFeedback} we obtain the diffusion-mediated Markovian feedback master equation
\begin{align}
	\hbar \, \dot{\rho} \equiv {\cal L}_{\rm mfb} \;\! \rho  
\label{DiffMediatedFBME}
	                         = {}& \! - \! i \big[\hat{H}_1,\rho \big] + {\cal D}[\hc]\rho + \hbar \;\! {\cal D}[\vop{f}]\rho  \nn \\
                               & - i \big\lceil \;\! \vop{f}, {\sf M}\dg \hc \;\! \rho + \rho \;\! {\sf M}\tp \hc\ddg \big\rfloor .
\end{align} 
Note that \eqref{DiffMediatedFBME} is valid for nonlinear systems (see Appendix~\ref{EngineeringDefintion}) as no assumptions about $\Hext$, $\hc$, and $\vop{f}$ were made in our derivation. That ${\cal L}_{\rm m}+{\cal L}_{\rm fb}$ is again of the Lindblad form is a rather lengthy exercise so we have proved it in Appendix~\ref{LindbladFBMasterEq}. The result may be written as 
\begin{align}
\label{FBMELindbladForm}  
 \hbar \, \dot{\rho} = {}& - i \big[ \Hext + \half ( \;\! \vop{f}\tp {\sf M}\dg \hc + \hc\dg {\sf M} \, \vop{f} \;\! ), \rho \big]  \nn \\
	                       & + {\cal D}\big[\hc-i{\sf M}\;\!\vop{f}\;\!\big] \rho 
	                         + {\cal D}\big[ \rt{\hbar \, \ID{R} - \MdgM} \, \vop{f} \;\!\big] \rho  \;,
\end{align}
where $\rt{\hbar \, \ID{R} - {\sf M}\dg {\sf M}}$ may be replaced by any matrix square root of $\hbar \, \ID{R} - {\sf M}\dg {\sf M}$ \footnote{In general the square root of a matrix $A$ is any matrix $B$ such that $B\dg B = A$. When $A \ge 0$, there exists a unique $B$ such that $B \ge 0$ and $B^2=A$, called the positive square root of $A$. The positive square root of $A$ is denoted by $\rt{A}$.}. We remark that while the Lindblad form is an important part of the theory, \eqref{FBMELindbladForm} is not necessarily more useful than \eqref{DiffMediatedFBME}. 
%
%due to the appearance of $\rt{\hbar\,\ID{R} - {\sf M}\dg {\sf M}}$. Given a diffusive measurement and its parameterization in terms of ${\sf M}$ one would have to calculate the (positive) matrix square root of $\hbar\;\!\ID{R} - {\sf M}\dg {\sf M}$ in order to obtain the master equation with feedback. On the other hand if one is content with \eqref{DiffMediatedFBME}, then the feedback master equation can be found directly once ${\sf M}$ is known.

\subsubsection{Feedback quantum Langevin equation}

Equation \eqref{DiffMediatedFBME}, or \eqref{FBMELindbladForm}, describes feedback in the \sch\ picture but they are not the only equations of motion capable of capturing the feedback process. An alternative theory of feedback exists in the \hei\ picture where feedback is described by a quantum Langevin equation for an arbitrary system vop $\vop{s}$. Such an equation follows unitary evolution and has the interpretation that measurements (namely the collapse of $\rho$ as occurs by using a measurement operator) never happens. Thus it also describes ``feedback without measurement'' \cite{Wis94}.

As before, the calculation can be simplified by first deriving the change in $\vop{s}$ due to feedback only and then adding it to the measurement contribution. This can be obtained from \eqref{QLEfb} by substituting in the expression for $\vop{y}_1$ and then normally ordering $\dBout$. The final result, including the measurement contribution is
\begin{widetext}
\begin{align}
\label{NonzeroDelayQLE}
	\hbar \, d\vop{s} = {}& \big( \, i \;\! [\hat{H}_{1},\vop{s}\;\!] 
	                        + {\cal J}[\hc\ddg] \;\! \vop{s} - \half \;\! \{ \hc\dg \hc, \vop{s} \} \, 
	                        + \hbar \;\! {\cal D}[\vop{f}] \;\! \vop{s} \;\! \big) dt 
	                        + [\;\!\hc\dg d\vop{B}_{\rm in} - d\vop{B}\dg_{\rm in} \hc, \vop{s}\;\!]  \nn \\
	                      & - i \mbrac{\vop{s}}{{\sf M}^{*} \,\vop{f}\;\!} \big[ \hc(t-\tau) dt + \dBin(t-\tau) \big] 
	                        + i \Big\{ \big[ \hc\dg(t-\tau) dt + \dBin\dg(t-\tau) \big] \mbrac{{\sf M} \,\vop{f}}{\vop{s}\;\!} \Big\}\tp  \nn \\
                        & - i \mbrac{\vop{s}}{\rt{\sf Z}^{\!*} \,\vop{f}\;\!} \dUin(t-\tau) 
                          + i \Big\{ \dUin\dg(t-\tau) \mbrac{\rt{\sf Z}\,\vop{f}}{\vop{s}\;\!} \Big\}\tp  \;.
\end{align}
\end{widetext}
The matrix $\rt{\sf Z}$ is the positive square root of \eqref{ZInTermsOfM} and $\dUin$ is an independent Wiener increment (recall \eqref{dUAppear1} and \eqref{dUAppear2}). One can check that \eqref{NonzeroDelayQLE} is a valid \ito\ equation, i.e.
\begin{align}
\label{DefnOfValidItoEqn}
	d\;\!(\vop{s} \;\! \hat{\alpha}) = (d\vop{s}) \;\! \hat{\alpha} + \vop{s} \, (d\hat{\alpha}) + (d\vop{s})(d\hat{\alpha})  \;,
\end{align}
for any operator $\hat{\alpha}$. Note that we can take the Markovian limit of \eqref{NonzeroDelayQLE} by setting $\tau=0$ in $\dBin(t-\tau)$ since $\bin(t)$ is continuous in time, although nowhere differentiable. The resulting equation with $\tau=0$ in \eqref{NonzeroDelayQLE} is then the \hei\-picure equivalent of \eqref{DiffMediatedFBME} in the sense that
\begin{align}
\label{SchHeiPicReln}
	d\an{\vop{s}} = {\rm Tr}\big[ \rho\smallsub{SB}(0) \, d\vop{s}(t) \big] = {\rm Tr}\smallsub{S}\big[ d\rho(t) \, \vop{s}(0) \big]  \;.
\end{align}
%

                                        %%%%%%%%%%%%%%%%%%%%%%%%%%%%%%%%%%%%%%%%%%%%%%%%%%%%%%%%%%%%%%%%%%%%%

\subsection{Diffusion-mediated Feedback Starting with Zero Feedback Delay}
\label{ZeroFeedbackDelayApproach}

When we allow the feedback delay to be zero we are letting the time at which $\vop{f}$ interacts with the bath converge to the same point in time as the interaction between $\hc$ and the bath. This eliminates the concept of $\bout$. Consequently the feedback interaction should be defined by 
\begin{align}
\label{ZeroDelayFeedbackHamiltonian}
	\Hfb = \hbar \: \vop{f}\tp \vop{y}_0  \;,
\end{align}
where $\vop{y}_0$ is
\begin{align}
\label{ZeroDelayCurrent}
	\hbar \, \vop{y}_0 \, dt = {\sf M}\dg \dBin + {\sf M}\tp \dBin\ddg + \hbar \, d\vog{\upsilon}_{\rm in}  \;. 
\end{align}
By working in the limit of zero feedback delay we are also allowing the measurement and feedback interactions to occur in the same infinitesimal time interval $[t,t+dt)$,
\begin{align}
\label{TotalUnitaryWithZeroDelay}
	\Umfb(t,t+dt) = \Ufb(t,t+dt) \, \Um(t,t+dt) \;.
\end{align}
where 
\begin{align}
	\Ufb(t,t+dt) = \exp\big(\!-i\Hfb dt/\hbar\big) = \exp\big(\!-i\,\vop{f}\tp\vop{y}_0 \, dt\big)  \;.
\end{align}
Since $\Hfb$ and $\Hm$ do not commute the order of $\Ufb$ and $\Um$ matters and the correct order is defined by the order in which the two processes happen in reality. This order should correspond to the order in which the unitaries act on a state, as shown in \eqref{TotalUnitaryWithZeroDelay}. That is the \sch\ picture is what defines the order in which we compose $\Um$ and $\Ufb$ to give $\Umfb$. When we evolve a vop $\vop{s}$ in the \hei\ picture from $t$ to $t+dt$ under measurement and feedback the order is then given by (with the unitary operators understood to act over an infinitesimal interval from $t$ to $t+dt$)
\begin{align}
\label{OrderOfUnitariesInHeiPic}
	\vop{s}(t+dt) = \Umfb\dg \, \vop{s} \, \Umfb = \Um\dg \, \Ufb\dg \, \vop{s} \, \Ufb \, \Um  \;.
\end{align}
There is of course nothing odd about letting $\Ufb$ act on $\vop{s}$ first in \eqref{OrderOfUnitariesInHeiPic}, it is simply a consequence of the definition of the \hei\ picture. 
If one insists on having $\Um$ act on $\vop{s}$ first, even in the \hei\ picture, then we can rewrite \eqref{OrderOfUnitariesInHeiPic} as
\begin{align}
	\vop{s}(t+dt) =  \hat{U}\dg_{\rm fb1} \Um\dg \, \vop{s} \, \Um \hat{U}_{\rm fb1}  \;,
\end{align}
where we have defined
\begin{align}
	\hat{U}_{\rm fb1} = \Um\dg \, \Ufb \, \Um = \exp\big(\!-\!i\, \hat{H}_{\rm fb1} \,dt / \hbar \big)  \;.
\end{align}
The Hamiltonian $\hat{H}_{\rm fb1}$ is given by 
\begin{align}
\label{InsistedOrderHamiltonian}
	\hat{H}_{\rm fb1} \, dt = \hbar \, \vop{f}\tp\!(t+dt) \, \vop{y}_1 \, dt \;,
\end{align}
and $\vop{y}_1$ is as before, given by \eqref{CurrentOperator}. Note that \eqref{InsistedOrderHamiltonian} has no ordering ambiguity (recall the discussion surrouding \eqref{requiredfbHamtau}) on its RHS since the current $\vop{y}_1$ appears at an (infinitesimally) earlier time than $\vop{f}(t+dt)$. In what follows we will take the former approach, i.e. with a vop in the \hei\ picture defined by \eqref{OrderOfUnitariesInHeiPic} and a feedback Hamiltonian given by \eqref{ZeroDelayFeedbackHamiltonian} and \eqref{ZeroDelayCurrent}.

The \hp\ equation for $\Umfb(t,t_{0})$ is 
\begin{widetext}
\begin{align}
\label{ZeroDelayHPE}
	\hbar \: d\Umfb(t,t_{0}) = {}& \Big[ \big( \!-\!i \Hext - \hc\dg \hc/2 - \hbar \, \vop{f}\tp \vop{f}/2 - i \, \vop{f}\tp {\sf M}\dg \hc \big) \, dt  
                                 + \dBin\dg \big( \hc - i \, {\sf M} \;\! \vop{f} \big) - \big( \hc\dg + i \, \vop{f}\tp {\sf M}\dg \big) \dBin  \nn \\
                               & - i \big( \vop{f}\tp \rt{\sf Z} \;\! \dUin + \dUin\dg \rt{\sf Z} \;\! \vop{f} \big) \Big] \Umfb(t,t_0)	\;,
\end{align}
with the initial condition $\Umfb(t_0,t_0)=\hat{1}$. From this we can derive yet another quantum Langevin equation,
\begin{align}
\label{ZeroFeedbackDelayQLE}
	\hbar \, d\vop{s} = {}& i \;\! \big[\hat{H}_{1},\vop{s}\;\!\big] \, dt + \hbar \, {\cal D}[\vop{f}] \;\! \vop{s} \: dt 
	                        - \mbrac{\vop{s}}{\hc\ddg} \big( \half \, \hc \;\! dt + \dBin \big)  
	                        - \Big\{ \big( \half \, \hc\dg dt + \dBin\dg \big) \mbrac{\hc}{\vop{s}} \Big\}\tp  \nn \\
	                      & + i \Big\{ \big( \hc\dg dt + \dBin\dg \big) \mbrac{{\sf M}\;\!\vop{f}}{\vop{s}} \Big\}\tp 
	                        - i \;\! \mbrac{\vop{s}}{{\sf M}^* \;\! \vop{f}\;\!} \big( \hc \;\! dt + \dBin \big)  
	                      	- i \;\! \mbrac{\vop{s}}{\rt{\sf Z}^* \;\!\vop{f}\;\!} \dUin  
                          + i  \Big\{ \dUin\dg \mbrac{\rt{\sf Z}\;\!\vop{f}}{\vop{s}\;\!} \Big\}\tp  \;.
\end{align}
\end{widetext}
This is again a valid \ito\ equation in the sense of \eqref{DefnOfValidItoEqn} and we have also placed the bath fields on the exterior so that terms containing $\dBin$ or $\dBin\dg$ vanish when averaged against a vacuum bath state.

%We said above (below \eqref{DefnOfValidItoEqn}) that setting $\tau=0$ in \eqref{NonzeroDelayQLE} gives us another valid \ito\ equation. 
Here we have to be careful that \eqref{ZeroFeedbackDelayQLE} is not quite the same as the equation which results from setting $\tau=0$ in \eqref{NonzeroDelayQLE}. Their difference lies in the last term (a commutator) in the first line of \eqref{NonzeroDelayQLE} and the last two terms in the first line of \eqref{ZeroFeedbackDelayQLE}. Let us illustrate the difference by first simplifying \eqref{NonzeroDelayQLE} and \eqref{ZeroFeedbackDelayQLE} by letting $\Hext=0$ and $\vop{f}={\bf 0}$ in both equations. In this case \eqref{NonzeroDelayQLE} simplifies to,
\begin{align}
\label{NonzeroDelayQLESimple}
	\hbar \, d\vop{s} = {}& \big( \, {\cal J}[\hc\ddg] \;\! \vop{s} - \half \;\! \{ \hc\dg \hc, \vop{s} \} \;\! \big) dt 
	                        + [\;\!\hc\dg d\vop{B}_{\rm in} - d\vop{B}\dg_{\rm in} \hc, \vop{s}\;\!]  \;,
\end{align}
while \eqref{ZeroFeedbackDelayQLE} simplifies to 
\begin{align}
\label{ZeroFeedbackDelayQLESimple}
	\hbar \, d\vop{s} = {}& - \mbrac{\vop{s}}{\hc\ddg} \big( \half \, \hc \;\! dt + \dBin \big)  \nn \\  
	                      & - \Big\{ \big( \half \, \hc\dg dt + \dBin\dg \big) \mbrac{\hc}{\vop{s}} \Big\}\tp \;.
\end{align}
To arrive at \eqref{ZeroFeedbackDelayQLESimple} we have used the fact that $\dBin$ will commute with an arbitrary system vop $\vop{s}$ at the same time. This means that in \eqref{ZeroFeedbackDelayQLESimple} (and also \eqref{ZeroFeedbackDelayQLE}) $\vop{s}$ is strictly a system vop; setting $\vop{s}$ to a bath vop in \eqref{ZeroFeedbackDelayQLE} would violate this assumption. Setting $\vop{s}=\bin$ in \eqref{ZeroFeedbackDelayQLE} (or \eqref{ZeroFeedbackDelayQLE}) yields the nonsensical result $\bout=\bin$. On the other hand \eqref{NonzeroDelayQLE}, and therefore \eqref{NonzeroDelayQLESimple}, uses only the commutability of $\dBout(t-\tau)$ with $\vop{s}(t)$ for $\tau > 0$. This is actually preserved even when we let $\vop{s}$ be a bath field. Indeed, when we set $\vop{s}=\bin$ in \eqref{NonzeroDelayQLESimple} (or \eqref{NonzeroDelayQLE}) the correct output relation of the bath field is obtained.

To derive a master equation we may move into the \sch\ picture from \eqref{ZeroDelayHPE} or simply take the average of \eqref{ZeroFeedbackDelayQLE}. Since \eqref{ZeroFeedbackDelayQLE} is normally ordered in the bath vops, its average with respect to \eqref{SysBathState} is simply 
\begin{align}
\label{AvgOfZeroFeedbackDelayQLE}
	\hbar \, \ban{d\vop{s}} = {}& \big\langle \;\! i \big[ \Hext,\vop{s} \big] 
	                              - \half \mbrac{\vop{s}}{\hc\ddg} \hc 
	                              - \half \big( \hc\dg \mbrac{\hc}{\vop{s}} \big)^{\!\top}  \nn \\
                              & + \hbar \, {\cal D}[\vop{f}] \;\!\vop{s}
                                + i \, \big( \hc\dg \mbrac{{\sf M}\;\!\vop{f}}{\vop{s}} \big)^{\!\top} 
                                - i \, \mbrac{\vop{s}}{{\sf M}^{*}\,\vop{f}\;\!} \;\! \big\rangle \, dt \;.
\end{align}
It is easy to show that
\begin{align}
	 - \half \mbrac{\vop{s}}{\hc\ddg} \hc - \half \big( \hc\dg \mbrac{\hc}{\vop{s}} \big)^{\!\top} 
	 = {}& \big( \hc\ddg \vop{s}\tp \big)^{\!\top} \hc - \half \big\{ \hc\dg \hc, \vop{s} \big\}  \nn \\
	 = {}& {\cal J}[\hc\ddg] \;\! \vop{s} - \half \big\{ \hc\dg \hc, \vop{s} \big\}  \;.
\end{align}
So the first line of \eqref{AvgOfZeroFeedbackDelayQLE} is just the average \eqref{QLEm} for which the contribution to $\dot{\rho}$ is well-known, given by \eqref{MasterEqnWithoutFeedback}. The first term on the second line is given by \eqref{FeedbackThree} while
\begin{align}
	\ban{\hc\dg \mbrac{{\sf M}\;\!\vop{f}}{\vop{s}}}\tp = {}& \Tr\Big\{ \vop{s} \;\! \big( \rho\smallsub{SB} \, \hc\dg {\sf M} \, \vop{f} 
	                                                          - \vop{f}\tp \rho\smallsub{SB} \;\! {\sf M}\tp \hc\ddg \big) \Big\}  \;,  \\
	\ban{\mbrac{\vop{s}}{{\sf M}^{*}\;\!\vop{f}}\;\!\hc} = {}& \Tr\Big\{ \vop{s} \;\! \big( \vop{f}\tp {\sf M}\dg \hc \;\! \rho\smallsub{SB} 
	                                                           - \hc\tp {\sf M}^* \rho\smallsub{SB} \;\! \vop{f} \big) \Big\} \;.
\end{align}
Therefore the second line of \eqref{AvgOfZeroFeedbackDelayQLE} is in fact \eqref{FeedbackQLE}. From these it should be clear that a master equation exactly of the form given by \eqref{DiffMediatedFBME} results, as expected. If we were not interested in the quantum Langevin equation then one would, and is in fact quicker, to derive the master equation directly from \eqref{ZeroDelayHPE}.

                                %%%%%%%%%%%%%%%%%%%%%%%%%%%   Diffusive Conditional Dynamics   %%%%%%%%%%%%%%%%%%%%%%%%%

\section{Conditional Dynamics}
\label{ConditionalEvolution}

\begin{figure}
\includegraphics[width=8.5cm]{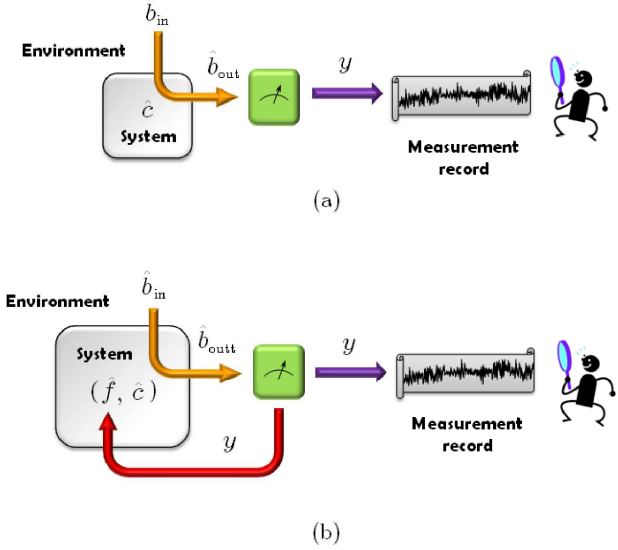} 
\caption{For simplicity we have shown only the SISO case with $L=1$. (a) The situation described by the works of Refs.~\cite{WD01,WD05}, in which the observer uses his/her knowledge of the measurement record $y$ (treated as a number in the \sch\ picture) to infer the state of the system but in the absence of feedback. (b) The situation described by the stochastic master equation \eqref{DiffusiveFeedbackSME}: The observer infers the state of the system from the measurement record in the presence of feedback. The inclusion of feedback (i.e. $\hat{f}$) changes the interaction between $\hat{b}_{\rm in}$ and the system. That is, $\hat{b}_{\rm in}$ now ``sees'' both $\hat{c}$ and $\hat{f}$, which is why we have written them as a pair in the figure. The result of this is to produce an output field $\hat{b}_{\rm outt}$ which evolves according to \eqref{ZeroDelayHPE}, which is different to $\hat{b}_{\rm out}$ (hence the extra ``t'' in the subscript). The actual system-bath input-output relation with respect to \eqref{ZeroDelayHPE} is worked out in Sec.~\ref{OutputCorrelations}.}
\label{ConditionalEvolutionPicture}
\end{figure}

To better understand applications of feedback one would like to know the controlled system dynamics as the monitoring and feedback occurs in real-time. It is well-known that continuously measured systems can be described by a nonlinear stochastic differential equation for the system state \cite{Bru02,JS06}. Here we will derive the a general diffusion-mediated stochastic feedback master equation in the \hei\ picture. This is an extension of the diffusive stochastic master equation found in Refs.~\cite{WD01,WD05} to include feedback but using a different parameterization of the measurement. We illustrate the two cases (with and without feedback) in Fig.~\ref{ConditionalEvolutionPicture}.

\subsection{Diffusion-mediated Feedback Stochastic Master Equation}

Previously we found the most general diffusive stochastic master equation with measurments alone to be given by,
\begin{align}
\label{DiffusiveSME}
	d\rhoc = {\cal L}_{\rm m} \;\! \rhoc \, dt + {\cal H}[d{\bf w}\tp {\sf M}\dg \hc] \;\!\rhoc  \;,
\end{align}
where ${\cal L}_{\rm m}$ is given by \eqref{MasterEqnWithoutFeedback} and $d{\bf w}$ is an $R \times 1$ (vector) Wiener increment defined by ${\rm E}[d{\bf w}(t)]={\bf 0}$ and
\begin{align}
	d{\bf w}(t) \, d{\bf w}\tp(t) = {}& \ID{R} \, dt  \;,  \\
	d{\bf w}(t) \, d{\bf w}\tp(t') = {}& 0  \quad  \forall \; t \ne t'  \;.
\end{align}
The superoperator ${\cal H}[\hat{A}]$, for any $\hat{A}$, is defined to be
\begin{align}
	{\cal H}[\hat{A}] \rho = \hat{A} \rho + \rho \hat{A}\dg - \bTr{\hat{A} \rho + \rho \hat{A}\dg} \rho  \;.
\end{align}
To generalize \eqref{DiffusiveSME} to account for feedback we first note that our foregoing derivation of the master equation prescribes us with the rule 
\begin{align}
\label{FeedbackRule1}
	{\cal L}_{\rm m}  \; \longrightarrow \; {\cal L}_{\rm mfb} = {\cal L}_{\rm m} + {\cal L}_{\rm fb} 
\end{align}
for the unconditioned evolution. But how does the conditional dynamics change? That is how can the nonlinear term in \eqref{DiffusiveSME} be altered to include feedback?

A derivation of the stochastic master equation in the \hei\ picture would be possible if we can establish a relation about the time evolution in the \sch\ and \hei\ pictures that involves the conditioning. For unconditional evolution such a relation is given by \eqref{SchHeiPicReln}, which made the derivation of the master equation possible in the \hei\ picture. We can in fact find an analogous relation that incorporates the conditioning of $\rho$ on the measured current. By considering the evolution over an infinitesimal time interval such an equation is given by
\begin{align}
\label{SchHeiPicRelnWithCond}
	   & \hbar^2 \, \Ban{\big( \ym - \an{\ym} \!\big) \, \vop{s}\tp\!(t+dt)} \, dt  \nn \\
	 {}& = \hbar^2 \, {\rm E}\!\;\!\Big\{ d{\bf w} \, {\rm Tr}\smallsub{S} \big[ \vop{s}\tp \rho_{{\bf y}_t}(t+dt) \big] \Big\}  \;,
\end{align}
where we have multiplied each side by $\hbar^2$ for convenience. This identity can be derived using quantum measurement theory. For simplicity we are assuming the state to be given at time $t$ (i.e. deterministic). The state on the RHS of \eqref{SchHeiPicRelnWithCond} is conditioned on the vector-valued current 
\begin{align}
	\hbar \, {\bf y} \, dt = \ban{{\sf M}\dg \hc + {\sf M}\dg \hc\ddg} \, dt + \hbar \, d{\bf w}  \;,
\end{align}
at only one time, $t$, where $d{\bf w}$ is a vector Wiener increment. To use \eqref{SchHeiPicRelnWithCond} we note from quantum measurement theory that 
any diffusive unravelling will be of the form
\begin{align}
\label{AnsatzSME}
	d\rhoc = {\cal L} \;\! \rhoc \, dt + {\cal H}[d{\bf w}\tp \vog{\alpha}\;\!] \;\! \rhoc \;,
\end{align}
for some $\vog{\alpha}$ and ${\cal L}$. We therefore make this ansatz in \eqref{SchHeiPicRelnWithCond} with $\vog{\alpha}$ to be determined by the LHS, which is in the \hei\ picture.

Using \eqref{AnsatzSME} and the fact that $\rhoc(t)\equiv\rho(t)$ is known, the RHS of \eqref{SchHeiPicRelnWithCond} simply reduces to 
\begin{align}
	  & \hbar^2 \, {\rm E}\!\;\!\Big\{ d{\bf w} \, {\rm Tr}\smallsub{S} \big[ \vop{s}\tp \rho_{{\bf y}_t}\!(t+dt) \big] \Big\}  \nn \\
	{}& =  \hbar \, \Tr \Big\{ d{\bf w} \, \vop{s}\tp {\cal H}[d{\bf w}\tp \vog{\alpha}\;\!] \;\! \rho \Big\}  \nn \\
\label{RHSFinal}
	{}& =  \hbar \, \Tr\Big\{ \big[ \;\! \vop{s} \,\big(\vog{\alpha}\tp \rho + \rho \;\! \vog{\alpha}\dg - \an{\vog{\alpha} 
	       + \vog{\alpha}\dg} \rho\;\!\big) \;\! \big]\tp \Big\} \: dt  \;.
\end{align}
The LHS of \eqref{SchHeiPicRelnWithCond} is 
\begin{align}
\label{LHSofSchHeiPicRelnWithCond}
	& \hbar^2 \,\big\langle ( \ym  - \an{\ym} ) \,\hbar\,\vop{s}\tp\!(t+dt) \;\!\big\rangle \, dt  \nn \\
	 {}& = \ban{(\hbar\,\ym dt) \,\hbar\,\vop{s}\tp\!(t+dt)} - \ban{\;\!\hbar\,\ym dt} \ban{\hbar\,\vop{s}\tp\!(t+dt)}  \;.
\end{align}
On substituting in \eqref{CurrentOperator}, the first term in \eqref{LHSofSchHeiPicRelnWithCond} is
\begin{align}
	\ban{(\hbar\,\ym dt) \:\hbar\,\vop{s}\tp\!(t+dt)} 
	= {}& \ban{{\sf M}\dg \dBmt\tp \,\hbar\,\vop{s}\tp\!(t+dt)\;\!}  \nn \\
	    & + \ban{{\sf M}\tp \dBmt\ddg \:\hbar\,\vop{s}\tp\!(t+dt)\;\!}  \nn \\
	    & + \ban{(\hbar \, \dupsin) \:\hbar\,\vop{s}\tp\!(t+dt)\;\!}  \;.
\end{align}  
By examining \eqref{ZeroFeedbackDelayQLE} it is not difficult to see that
\begin{align}
\label{ConditionalTerm1}
	& \ban{{\sf M}\dg \dBmt\tp \:\hbar\,\vop{s}\tp\!(t+dt)\;\!} \nn \\
	{}& = \hbar \, \ban{\;\![\;\!\vop{s}\,({\sf M}\dg \hc)\tp]\tp} \, dt  
	      +  i \;\! \hbar \, \ban{\mbrac{{\sf M}\dg{\sf M}\;\!\vop{f}}{\vop{s}\;\!}} \, dt  \;, \\[0.2cm]
\label{ConditionalTerm2}
	& \ban{{\sf M}\tp \dBmt\ddg \:\hbar\,\vop{s}\tp\!(t+dt)\;\!} = \hbar \, \an{{\sf M}\tp \hc\ddg \vop{s}\tp} \, dt  \;, \\[0.2cm]
\label{ConditionalTerm3}
	& \ban{(\hbar \,\dupsin)\:\hbar \,\vop{s}\tp\!(t+dt)} \nn \\
	{}& = i \;\! \hbar^2 \, \ban{\mbrac{\vop{f}}{\vop{s}}} \, dt 
	      - i \;\! \hbar \, \ban{\mbrac{{\sf M}\dg{\sf M}\;\!\vop{f}}{\vop{s}\;\!}} \, dt \;,  \\[0.2cm]
\label{ConditionalTerm4}
	& \ban{\;\!\hbar\,\ym dt} \ban{\;\!\hbar\,\vop{s}\tp\!(t+dt)}  \nn \\
	{}& = \hbar \an{{\sf M}\dg \hc} \an{\vop{s}\tp} dt + \hbar \an{{\sf M}\tp \hc\ddg} \an{\vop{s}\tp} dt \;.
\end{align}
Writing \eqref{ConditionalTerm1}--\eqref{ConditionalTerm4} as a trace we get
\begin{align}
\label{LHSFinal}
	& \hbar^2 \,\big\langle \big( \ym  - \an{\ym} \!\big) [\;\!\hbar\,\vop{s}\tp\!(t+dt)] \;\!\big\rangle \, dt  \nn \\[0.2cm]
	{}& = \hbar \,\Tr\Big\{ \big[ \;\! \vop{s} \;\! \{ ({\sf M}\dg \hc - i \hbar \, \vop{f})\tp \rho 
	      + \rho \;\! ({\sf M}\dg \hc - i \hbar \, \vop{f})\dg  \nn \\
	   & \hspace{0.3cm} - \an{({\sf M}\dg \hc)\tp + ({\sf M}\dg \hc)\dg} \rho \;\!\} \;\!\big]\tp \Big\} \,dt \;.
\end{align}
Equating \eqref{LHSFinal} to \eqref{RHSFinal} and solving for $\vog{\alpha}$ gives
\begin{align}
\label{AlphaWithFeedback}
	\vog{\alpha} = {\sf M}\dg \hc - i \;\! \hbar \, \vop{f} \;.
\end{align}
Invoking \eqref{FeedbackRule1} and \eqref{AlphaWithFeedback}, we arrive at the diffusion-mediated stochastic feedback master equation 
\begin{align}
\label{DiffusiveFeedbackSME}
	\hbar \: d\rhoc = {}& \! - \! i \big[\hat{H}_1,\rhoc \big] + {\cal D}[\hc]\rhoc + \hbar \, {\cal D}[\vop{f}]\rhoc  \nn \\
                      & - i \sbrac{\;\!\vop{f}}{{\sf M}\dg \hc\;\!\rhoc + \rhoc\;\!{\sf M}\tp \hc\ddg}  \nn \\
                      & + {\cal H}\big[d{\bf w}\tp ({\sf M}\dg \hc - i \hbar \, \vop{f}\;\!) \big] \;\! \rhoc  \;.
\end{align}
Comparing \eqref{DiffusiveFeedbackSME} to \eqref{DiffusiveSME} we can summarize the changes necessary to include feedback in the stochastic master equation \eqref{DiffusiveSME} by the two transformations
\begin{align}
\label{FeedbackRule1SecondTime}
	\Lm \; \longrightarrow \; & \Lmfb  \;, \\
\label{FeedbackRule2}
	\mrep\dg \hc \; \longrightarrow \; & \mrep\dg \hc - i \hbar \, \vop{f}  \;.
\end{align}

We can understand why $\vop{f}$ must appear in the nonlinear term by considering the case when ${\sf M}=0$. In this case the feedback master equation is simply
\begin{align}
\label{ZeroMFeedbackMasterEqn}
	\hbar \, \dot{\rho} = -i\;\![\Hext,\rho\;\!] + {\cal D}[\hc] \rho + \hbar \, {\cal D}[\vop{f}] \rho  \;,
\end{align}
and the current fed back is pure noise
\begin{align}
\label{ZeroMCurrent1}
	{\bf y} \, dt = d{\bf w}  \;.
\end{align}
Equation \eqref{ZeroMFeedbackMasterEqn} is the unconditional evolution for the measurement defined by \eqref{ZeroMCurrent1}. If we now condition the state on the pure-noise output then the stochastic master equation which unravels \eqref{ZeroMFeedbackMasterEqn} is
\begin{align}
\label{ZeroMStochasticFeedbackMasterEqn}
	\hbar \, d\rhoc = {}& -i\;\![\Hext,\rhoc] + {\cal D}[\hc] \rhoc + \hbar \, {\cal D}[\vop{f}] \rhoc \nn \\ 
	                    & + \hbar \, {\cal H}[-i\;\!d{\bf w}\tp\vop{f}\;\!] \rhoc
\end{align}
This can be seen by noting that \eqref{ZeroMCurrent1} can also be written as
\begin{align}
\label{ZeroMCurrent2}
	{\bf y} \, dt = \ban{-i\;\!\vop{f} + (-i\;\!\vop{f})\ddg} + d{\bf w}   \;,
\end{align}
which gives rise to the nonlinear term in \eqref{ZeroMStochasticFeedbackMasterEqn}. When ${\sf M}\ne 0$ we get the general case of \eqref{DiffusiveFeedbackSME}.

\subsection{Output Correlation Function}
\label{OutputCorrelations}

When we include feedback in our theory the controlled dynamics can be accounted for by transforming the input fields according to $\Umfb$ as opposed to $\Um$. That is, instead of \eqref{MmtOutputField} we now have the new output field
\begin{align}
\label{dBmfb}
	\dBmfb(t) \equiv {}& \Umfb\dg(t+dt,t) \, \dBin(t) \, \Umfb(t+dt,t)  \nn \\
            = {}& \big[ \hc(t)  - i \;\! {\sf M} \;\! \vop{f}(t) \;\! \big] \, dt  + \dBin \;,
\end{align}
which can be derived from \eqref{ZeroDelayHPE}. The use of the subscript ``outt'' is deliberate, to be read as ``out twice''. This is to remind us that $\dBmfb$ is the output field obtained from using $\Umfb$, which is a composition of two unitaries \footnote{Note however that $\Umfb\dg \bin \Umfb \ne \Ufb\dg \bout \Ufb$.}. The input field $\dBin$ would still have the same mop-bracket with an arbitrary system vop $\vop{s}$ as given by \eqref{InputSysCommutator}, but $\dBmt$ in \eqref{OutputSysCommutator} should be replaced by $\dBmfb$. Thus we now have
\begin{align}
\label{InputSysCommutatorWithFeedback}
	\big\lfloor \dBin(t), \vop{s}(t') \big \rceil = 0 \quad \forall \; t' \le t \;, \\
\label{OutputSysCommutatorWithFeedback}
	\big\lfloor \dBmfb(t), \vop{s}(t') \big\rceil = 0 \quad \forall \; t' > t  \;.
\end{align} 
We should not forget to change the input field $\dUin$ as well since it will now evolve under the dynamics of feedback. Recall that $\dUin$ was introduced in \eqref{dUAppear1}, where it appeared as a vacuum noise in the current that did not interact with the system. When we add feedback this noise is redirected onto the system so it is no longer correct to assume that it is independent of the system as was the case in \eqref{dUAppear1}. We thus have an additional input-output relation, which can also be derived from \eqref{ZeroDelayHPE},
\begin{align}
\label{dUmfb}
	\dUmfb(t) \equiv {}& \Umfb\dg(t+dt,t) \, \dUin(t) \, \Umfb(t+dt,t)  \nn \\
	               = {}& \dUin(t) - i \;\! \rt{\sf Z} \, \vop{f}(t) \, dt  \;.
\end{align}
Similarly to \eqref{InputSysCommutatorWithFeedback} and \eqref{OutputSysCommutatorWithFeedback},
\begin{align}
\label{dUinSysCommutatorWithFeedback}
	\big\lfloor \dUin(t), \vop{s}(t') \big \rceil = 0 \quad \forall \; t' \le t \;, \\
\label{dUmfbSysCommutatorWithFeedback}
	\big\lfloor \dUmfb(t), \vop{s}(t') \big\rceil = 0 \quad \forall \; t' > t  \;.
\end{align} 
Relations \eqref{dBmfb} and \eqref{dUmfb} in turn define a new vop-valued current
\begin{align}
\label{ymfb}
	\hbar \, \ymfb \, dt = {\sf M}\dg \dBmfb + {\sf M}\tp \dBmfb\ddg + \hbar \, \dupsmfb  \;,
\end{align}
where the subscript on the current should remind us that it is defined in terms of $\dBmfb$, or the number of times the letter ``t'' appears on the RHS.
\begin{align}
\label{dupsmfb}
	\hbar \, \dupsmfb \equiv {}& \rt{\sf Z} \, \dUmfb + \rt{\sf Z}^{*} \, \dUmfb\ddg   \nn \\
	                       = {}& \hbar \, d\vog{\upsilon}_{\rm in} - i \;\! {\sf Z} \, \vop{f} \, dt
	                             + i \;\! {\sf Z}^* \, \vop{f} \, dt  \;.
\end{align}
From \eqref{dBmfb}, \eqref{ymfb}, and \eqref{dupsmfb}, we can see that
\begin{align}
\label{CurrentEquivalence}
 	\ymfb = \ym  \;.
\end{align} 
That is the current evolved over an infinitesimal interval from $t$ to $t+dt$ under both measurement and feedback is in fact the same as the current evolved in the same time interval but with measurement alone. Equation \eqref{CurrentEquivalence} can also be seen from the form of the Hamiltonian \eqref{ZeroDelayFeedbackHamiltonian}, which gives
\begin{align}
	\big[ \Hfb,\vop{y}_0 \big] = 0  \;.
\end{align}
Substituting $\Umfb$ into the definition of $\ymfb$ we obtain   
\begin{align}
	\ymfb(t) = {}& \Umfb\dg(t+dt,t) \, \vop{y}_{0}(t) \, \Umfb(t+dt,t) \nn \\
	         = {}& \Um\dg(t+dt,t) \, \vop{y}_{0}(t) \, \Um(t+dt,t) = \ym(t)  \;.
\end{align}
Using \eqref{ymfb} we can calculate how the current at time $t$ is correlated to the current at a later time $t+\tau$ during which feedback is applied. The time separation $\tau$ is assumed to be non-negative. We then obtain 
\begin{widetext}
\begin{align}
\label{CorrFuncnWithFeedback1}
	\hbar^2 \ban{\ymfb(t) \, \ymfb\tp(t+\tau)} 
	= {}& \ban{{\sf M}\dg \;\! \bmfb(t) \, \bmfb\tp(t+\tau) \, {\sf M}^*}  
	      + \ban{{\sf M}\dg \;\! \bmfb(t) \, \bmfb\dg(t+\tau) \, {\sf M}}
	      + \ban{{\sf M}\dg \;\! \bmfb(t) \, \hbar \;\! \zetamfb\tp(t+\tau)}  \nn \\
	    & + \ban{{\sf M}\tp \;\! \bmfb\ddg(t) \, \bmfb\tp(t+\tau) \, {\sf M}^*}
	      + \ban{{\sf M}\tp \;\! \bmfb\ddg(t) \, \bmfb\dg(t+\tau) \, {\sf M}} 
	      + \ban{{\sf M}\tp \;\! \bmfb\ddg(t) \, \hbar \;\! \zetamfb\tp(t+\tau)}  \nn \\
      & + \ban{\hbar \, \zetamfb(t) \, \bmfb\tp(t+\tau) \, {\sf M}^*}  
        + \ban{\hbar \, \zetamfb(t) \, \bmfb\dg(t+\tau) \, {\sf M}}
        + \ban{\hbar \, \zetamfb(t) \, \hbar \, \zetamfb\tp(t+\tau)}  \,.
\end{align}
\end{widetext}
%
%
%
%
%\begin{widetext}
%\begin{align}
%\label{CorrFuncnWithFeedback1}
%	\hbar^2 \ban{\ymfb(t) \, \ymfb\tp(t+\tau)} 
%	
%	= {}& \ban{\,[ \;\!{\sf M}\dg \bmfb(t) + {\sf M}\tp \bmfb\ddg(t) + \hbar \, \zetamfb(t) ]  
%	             [ \;\! \bmfb\tp(t+\tau) \;\! {\sf M}^* + \bmfb\dg(t+\tau) \;\! {\sf M} + \hbar \, \zetamfb\tp(t+\tau) ] }  \nn \\[0.2cm]
%
%
%	= {}& \underset{\text{term A}}{\ban{{\sf M}\dg \;\! \bmfb(t) \, \bmfb\tp(t+\tau) \, {\sf M}^*}}  
%	      + \underset{\text{term B}}{\ban{{\sf M}\dg \;\! \bmfb(t) \, \bmfb\dg(t+\tau) \, {\sf M}}}
%	      + \underset{\text{term E}}{\ban{{\sf M}\dg \;\! \bmfb(t) \, \hbar \;\! \zetamfb\tp(t+\tau)}}  \nn \\
%%	      
%	    & + \underset{\text{term C}}{\ban{{\sf M}\tp \;\! \bmfb\ddg(t) \, \bmfb\tp(t+\tau) \, {\sf M}^*}}
%	      + \underset{\text{term D}}{\ban{{\sf M}\tp \;\! \bmfb\ddg(t) \, \bmfb\dg(t+\tau) \, {\sf M}}} 
%	      + \underset{\text{term F}}{\ban{{\sf M}\tp \;\! \bmfb\ddg(t) \, \hbar \;\! \zetamfb\tp(t+\tau)}}  \nn \\
%%	      
%      & + \underset{\text{term G}}{\ban{\hbar \, \zetamfb(t) \, \bmfb\tp(t+\tau) \, {\sf M}^*}}  
%        + \underset{\text{term H}}{\ban{\hbar \, \zetamfb(t) \, \bmfb\dg(t+\tau) \, {\sf M}}}
%        + \underset{\text{term I}}{\ban{\hbar \, \zetamfb(t) \, \hbar \, \zetamfb\tp(t+\tau) }} \,.
%\end{align}
%\end{widetext}
%
%
%
Note that here we have introduced quantum stochastic processes $\zetamfb$ and $\mumfb$, defined in terms of the increments by
\begin{align}
\label{ZetaProcessDefn}
	\dupsmfb = \zetamfb\,dt = \big( \rt{\sf Z} \mumfb + \rt{\sf Z}^* \mumfb\ddg \big) dt \;,
\end{align}
where
\begin{align}
\label{MuProcessDefn}
	\dUmfb = \mumfb \, dt = \big( \muin - i \rt{\sf Z} \;\! \vop{f} \;\! \big) dt  \;.
\end{align}
As with earlier calculations, the assumption of a vacuum bath state suggests that we should substitute \eqref{ZetaProcessDefn} into \eqref{CorrFuncnWithFeedback1} and then normal and time order each term before the average is taken. The output field vops satisfy the familiar free-field mop-brackets
\begin{align}
	\mbrac{\bmfb(t)}{\bmfb\ddg(t')} = \hbar \: \ID{L} \, \delta(t-t')  \quad  \forall \; t,t'\;,  
\end{align}
and also
\begin{align}
	\mbrac{\bmfb(t)}{\bmfb(t')} = \mbrac{\bmfb\ddg(t)}{\bmfb\ddg(t')} = {}& 0  \quad  \forall \; t,t' \;.
\end{align}
The same is true for $\mumfb$ since it is also a free field, but remember that $\mumfb\,$ is $R\times 1$ so 
\begin{align}
	\mbrac{\mumfb(t)}{\mumfb\ddg(t')} = \hbar \: \ID{R} \, \delta(t-t') \;.  
\end{align}
We also have, and it is not difficult to see, that
\begin{align}
	\mbrac{\bmfb(t)}{\mumfb(t')} = \mbrac{\bmfb(t)}{\mumfb\ddg(t')} = 0  \quad \forall \; t,t'  \;.
\end{align}
\begin{widetext}
We summarize the result of each term in Appendix~\ref{DerivanOfCorrFuncnWithFeedback}. Using the results therein we arrive at
\begin{align}
\label{CorrFuncnWithFeedback2}
	\hbar^2 \ban{\ymfb(t) \, \ymfb\tp(t+\tau)} 
	= {}& \ban{[\;\!{\sf M}\dg \hc(t+\tau) + {\sf M}\tp \hc\ddg(t+\tau)] [\;\!\hc\tp\!(t) \;\! {\sf M}^{*} - i \;\!\hbar \;\!\vop{f}\tp\!(t)]}\tp  \nn \\
	    & + \ban{[\;\!{\sf M}\tp \hc\ddg(t) + i\;\!\hbar \;\! \vop{f}(t)] [\;\!\hc\tp\!(t+\tau) \;\! {\sf M}^{*} + \hc\dg(t+\tau) {\sf M}]} 
	      + \hbar^2 \;\! \ID{R} \, \delta(\tau)  \;.
\end{align}
Applying vop quantum regression formulas to \eqref{CorrFuncnWithFeedback2} the final result is 
\begin{align}
\label{CorrFuncnWithFeedback3}
	\hbar^2 \ban{\ymfb(t) \, \ymfb\tp(t+\tau)} 
	= \Bigg( \Tr \Big\{ \big( {\sf M}\dg \hc + {\sf M}\tp \hc\ddg \big) 
	  e^{{\cal L}_{\rm mfb}} \Big[ \big(\hc\tp {\sf M}^* - i\;\!\hbar\;\!\vop{f}\tp \big) \rho(t) 
	  + \rho(t) \big( \hc\dg {\sf M} + i\;\!\hbar\;\!\vop{f}\tp \big) \Big] \Big\} \Bigg)^{\!\!\top} + \, \hbar^2 \;\! \ID{R} \, \delta(\tau) \;,
\end{align}
\end{widetext}
where the time-dependence has been placed in the system state and the vops are time-independent. Note that \eqref{CorrFuncnWithFeedback3} could have obtained by using the transformations \eqref{FeedbackRule1SecondTime} and \eqref{FeedbackRule2} in the measurement-only correlation function
\begin{align}
\label{MmtOnlyCorrFunction}
	{}& \hspace{-0.5cm} \hbar^2 \, \ban{\vop{y}_{1}(t) \, \vop{y}\tp_{1}(t+\tau)}  \nn \\
	   = {}& \left( {\rm Tr}\Big\{ \big(\mrep\dg \hc + \mrep\tp \hc\ddg \big) 
	      e^{\Lm\tau} \big[ \hc\tp \mrep^* \, \rho(t) + \rho(t) \, \hc\dg \mrep \big] \Big\} \right)\tp  \nn \\
	    {}& + \, \hbar^2 \, \ID{R} \, \delta(\tau) \;.
\end{align}
as one might have guessed.
%
%We first generalize the autocorrelation of the homodyne current found previously in Ref.~\cite{WM93b} without feedback to the case of $L>1$. Written in our notation it reads
%%
%\begin{align}
%	{}& \ban{\vop{y}_{1}(t) \, \vop{y}\tp_{1}(t+\tau)}  \nn \\
%	  & = \Big( {\rm Tr}\Big\{ \big(\hc+\hc\dg\big) e^{\Lm\tau} \big[ \hc \, \rho(t) + \rho(t) \, \hc\dg \big] \Big\} \Big)\tp  \;.
%\end{align}
%

                               %%%%%%%%%%%%%%%%%%%%%%%%%%%%%  Consistency with Previous Results  %%%%%%%%%%%%%%%%%%%%%%%%%%%%%%%

\section{Simple Cases}
\label{ConsistencyWithPrevResults}

Here we illustrate how the above theory can be used by considering Markovian feedback mediated by homodyne and heterodyne detection. For simplicity we take $L=1$. In the case of homodyne-mediated feedback we then obtain a SISO theory whereas for heterodyne-mediated feedback we get a one-input two-output theory. In the case of homodyne-mediated feedback we recover results previously derived in Refs.~\cite{Wis94} and \cite{WM94}. We will allow for non-unit detection efficiency in both cases and write $\eta$ in place of ${\sf H}$.

%show how to recover previous results derived in Refs.~\cite{Wis94} and \cite{WM94a} as the limit of the above theory. These include feedback master equations mediated by a homodyne and heterodyne measurement of the output field $\hat{b}_{\rm m}$. We will allow for non-unit detection efficiency $\eta\equiv{\sf H}$. 

\subsection{Consistency with Previous Results --- Homodyne-mediated feedback}

Consider the SISO limit defined by a quadrature measurement of the form
\begin{align}
\label{HomX}
	\hbar \, \ban{\hat{y}_1} \, dt \propto  \ban{d\hat{B}_{\rm out} + d\hat{B}\dg_{\rm out}}  \;.
\end{align}
The condition that ${\sf M}$, now a scalar, must satisfy is simply
\begin{align}
	|{\sf M}|^2 = \hbar \, \eta  \;.
\end{align}
The measurement defined by \eqref{HomX} can be achieved by choosing ${\sf M}=\rt{\hbar\,\eta}$. This gives ${\sf Z}=\hbar \;\!\bar{\eta}$, where we have defined $\bar{\eta}=1-\eta$ for convenience. We find
\begin{align}
\label{HomodyneCurrent}
	\hbar \, \hat{y}_{\rm 1} \;\! dt = \rt{\hbar\,\eta} \big( \hat{c} + \hat{c}\dg \big) dt + \hbar \, d\hat{v}_{\rm m}  \;,
\end{align}
where the measurement noise is 
\begin{align}
	\hbar \, d\hat{v}_{\rm m} = 2 \, \rt{\hbar\,\eta} \, \Re[d\hat{B}_{\rm in}] + 2 \rt{\hbar\;\!\bar{\eta}} \, \Re[d\hat{U}_{\rm in}]  \;.
\end{align}
It is clear that
\begin{align}
	\big( \hbar \, d\hat{v}_{\rm m} \big)^2 
	= 4 \;\! \hbar \;\! \eta \, \big(\Re[d\hat{B}_{\rm in}]\big)^2 + 4 \;\! \hbar \;\! \bar{\eta} \, \big(\Re[d\hat{U}_{\rm in}]\big)^2 = \hbar^2 dt \;.
\end{align}
From \eqref{DiffusiveFeedbackSME} the stochastic feedback master equation is then
\begin{widetext}
\begin{align}
\label{SISODiffusiveFBSME}
	\hbar \;\! d\rhoc = \Big( \!- i\;\![\Hext,\rhoc]  + {\cal D}[\hat{c}] \rhoc  + \hbar \;\! {\cal D}[\hat{f}] \rhoc 
	                    - i \rt{\hbar\;\!\eta}\;\![\hat{f}, \hat{c}\;\!\rhoc + \rhoc \;\!\hat{c}\dg \;\!]  \Big) \, dt 
	                    + dw \, {\cal H}[\rt{\hbar\;\!\eta} \hat{c} - i \hbar \hat{f}\;\!] \rhoc\;.
\end{align}
This is consistent with the stochastic master equation found in Ref.~\cite{WM94} for $\hbar=1$ and when the current is suitably rescaled. It also reproduces the master equation in Ref.~\cite{Wis94} when $\hbar=\eta=1$. The Lindblad form of the unconditioned evolution can be found directly from \eqref{FBMELindbladForm},
\begin{align}
\label{SISODiffusiveFBME}
	\hbar \;\! \dot{\rho} \equiv {\cal L}_{\rm hom} \, \rho
	                           = - i \;\! [\Hext + \half \rt{\hbar\;\!\eta} (\hat{f} \hat{c} + \hat{c}\dg \hat{f}), \rho] 
	                             + {\cal D}[\hat{c}-i\rt{\hbar\;\!\eta}\hat{f}\;\!] \rho 
	                             + \hbar \;\! \bar{\eta} \;\! {\cal D}[\hat{f}\;\!] \rho  \;.
\end{align}
Again, this is consistent with the Lindblad form obtained in Ref.~\cite{WM94} (for $\hbar=1$) and Ref.~\cite{Wis94} (for $\hbar=\eta=1$), but in these works the Lindblad form was obtained by algebraic manipulation of \eqref{SISODiffusiveFBSME}. The two-time correlation function of \eqref{HomodyneCurrent} is, from \eqref{CorrFuncnWithFeedback3},
\begin{align}
	\hbar^2 \ban{\hat{y}_2(t) \, \hat{y}_2(t+\tau)}
	= \rt{\hbar \;\! \eta} \, \Tr \Big\{ \big( \hat{c} + \hat{c}\dg \big) 
	  e^{{\cal L}_{\rm hom}\;\!\tau} \big[ ( \rt{\hbar\;\!\eta} \hat{c} - i \hbar \hat{f} ) \rho 
	  + {\rm Hc} \big] \Big\}  +  \hbar^2 \, \delta(\tau)  \;.
\end{align}
\end{widetext}
and reproduces (4.10) of Ref.~\cite{Wis94} when $\hbar=\eta=1$.

We can also find a Markovian quantum Langevin equation from either \eqref{NonzeroDelayQLE} or \eqref{ZeroFeedbackDelayQLE}. There is no restriction on the number of components that $\vop{s}$ is allowed. For simplicity we take it to be a scalar-operator. Taking the Markovian limit of \eqref{NonzeroDelayQLE} the homodyne feedback quantum Langevin equation is
\begin{align}
\label{HomodyneQLE}
	\hbar \, d\hat{s} = {}& \big( \, i \;\! [\hat{H}_{1},\hat{s}\;\!] 
	                        + {\cal J}[\hat{c}\dg] \;\! \hat{s} - \half \;\! \{ \hat{c}\dg \hat{c}, \hat{s} \} \, 
	                        + \hbar \, {\cal D}[\hat{f}] \;\! \hat{s} \;\! \big) dt   \nn \\
	                      & + \big[\;\!\hat{c}\dg d\hat{B}_{\rm in} - d\hat{B}\dg_{\rm in} \hat{c}, \hat{s}\;\!\big]  \nn \\
	                      & - i \,\rt{\hbar\;\!\eta} \big[\hat{s},\hat{f}\;\!\big] \big( \hat{c} \, dt + d\hat{B}_{\rm in} \big)  \nn \\
	                      & + i \,\rt{\hbar\;\!\eta} \big( \hat{c}\dg \, dt + d\hat{B}\dg_{\rm in} \big) \big[ \hat{f},\hat{s}\;\!\big]  \nn \\
                        & - i \rt{\hbar\;\!\bar{\eta}} \big[\hat{s},\hat{f}\;\!\big] d\hat{U}_{\rm in}  
                          + i \rt{\hbar\;\!\bar{\eta}} \, d\hat{U}\dg_{\rm in} \big[\hat{f},\hat{s}\;\!\big]  \;. 
\end{align}
As before, when $\hbar=\eta=1$ this correctly reproduces (4.16) of Ref.~\cite{Wis94}. Note the extra noise terms $d\hat{U}_{\rm in}$ and $d\hat{U}\dg_{\rm in}$ in \eqref{HomodyneQLE} which do not appear in (4.16) of Ref.~\cite{Wis94}, since there, the quantum Langevin equation was derived in the limit of $\eta=1$.

\subsection{Heterodyne-mediated feedback}

A heterodyne detection is equivalent to two homodyne measurements of orthogonal quadratures each with half the detection efficiency so this requires $R=2$. Consider the heterodyne current defined by
\begin{align}
	\hbar \, \ban{\ym} \, dt  \, \propto \, \rt{\frac{\eta}{2}} 
	                                        \tbo{\ban{d\hat{B}_{\rm out} + d\hat{B}\dg_{\rm out}}}{-i\ban{d\hat{B}_{\rm out} - d\hat{B}\dg_{\rm out}}} \;.
\end{align}
This can be effected by 
\begin{align}
\label{HeterodyneM}
	{\sf M} = \rt{\frac{\hbar\;\!\eta}{2}} \obt{1}{i}  \;,
\end{align}
which satisfies \eqref{DefnOfM}. 
%
%\begin{align}
%	{\sf M}\dg \hc \rho + \rho \;\! {\sf M}\tp \hc\ddg 
%	= \rt{\frac{\hbar\;\!\eta}{2}} \tbo{\hat{c} \rho + \rho \hat{c}\dg}{-i(\hat{c} \rho - \rho \hat{c}\dg)} \;.
%\end{align}
%
%
The stochastic master equation from \eqref{DiffusiveFeedbackSME} is thus
\begin{widetext}
\begin{align}
\label{HeterodyneFeedbackSME}
	\hbar \, d\rhoc = {}& \Big( \!-i\big[\Hext,\rhoc\big] + {\cal D}[\hat{c}] \rhoc + \hbar\,{\cal D}[\hat{f}_1] \rhoc 
	                      + \hbar\, {\cal D}[\hat{f}_2] \rhoc  
	                      - i \, \rt{\frac{\hbar\;\!\eta}{2}} \big[\hat{f}_1, \hat{c} \rhoc + \rhoc \hat{c}\dg\big]   
	                      - i \, \rt{\frac{\hbar\;\!\eta}{2}} \big[\hat{f}_2, -i(\hat{c} \rhoc - \rhoc \hat{c}\dg)\big] \Big) dt  \nn \\
                      & + dw_1 \, {\cal H}[\rt{\hbar\;\!\eta/2} \, \hat{c} - i \hbar \hat{f}_1\;\!] \rhoc  
                        + dw_2 \, {\cal H}[-i\rt{\hbar\;\!\eta/2} \, \hat{c} - i \hbar \hat{f}_2\;\!] \rhoc  \;.
\end{align}
\end{widetext}
Setting $\hbar=\eta=1$ this is consistent with a special case of the heterodyne feedback master equation Ref.~\cite{WM94a} (see (5.19)--(5.24) with $N=M=0$). The Lindblad form of the master equation unravelled by \eqref{HeterodyneFeedbackSME} can be obtained by noting that \eqref{HeterodyneM} leads to 
\begin{align}
	{\sf Z} = \hbar \tbt{1-\eta/2}{-i\;\!\eta/2}{i\;\!\eta/2}{1-\eta/2} \;,
\end{align}
which has the positive square root
\begin{align}
	\rt{\sf Z} = \frac{\rt{\hbar}}{2} \tbt{1+\rt{\bar{\eta}}}{-i\;\!(1-\rt{\bar{\eta}})}{i(1-\rt{\bar{\eta}})}{1+\rt{\bar{\eta}}}  \;.
\end{align}
By introducing 
\begin{align}
	\hat{F} = \hat{f}_1 + i \hat{f}_2  \;,
\end{align}
the unconditioned evolution unravelled by \eqref{HeterodyneFeedbackSME} has a Lindblad form which can be written compactly as
\begin{align}
	\hbar \, \dot{\rho} \equiv {\cal L}_{\rm het} \, \rho
	                         = {}& - i \big[ \Hext,\rho \big] 
	                               - i \, \rt{\frac{\hbar\eta}{8}} \, \big[\hat{F}\dg \hat{c} + \hat{c}\dg \hat{F}, \rho \big]  \nn \\
	                             & + \frac{\hbar}{4} \, {\cal D}\big[\hat{F}\dg + \rt{\bar{\eta}} \hat{F}\big] \rho
	                               + \frac{\hbar}{4} \, {\cal D}\big[\hat{F}\dg - \rt{\bar{\eta}} \hat{F}\big] \rho  \nn \\
	                             & + {\cal D}\big[\hat{c} - i \rt{\frac{\hbar \eta}{2}} \hat{F}\big] \rho  \;.
\end{align}
We find the heterodyne current has correlations given by
\begin{widetext}
\begin{align}
	& \hbar^2 \, \ban{\ymfb(t) \, \ymfb\tp(t+\tau)}  \nn \\
	{}& = \rt{\frac{\hbar\;\!\eta}{2}} 
 	\tbt{\Tr \Big\{ \big(\hat{c}+\hat{c}\dg\big) \;\! 
 	     e^{{\cal L}_{\rm het}\;\!\tau} \Big[ \Big( \rt{\frac{\hbar\;\!\eta}{2}} \hat{c}- i \hbar \hat{f}_1 \Big) \rho + {\rm Hc} \Big] \Big\}}
	    {\Tr \Big\{ -i\big(\hat{c}-\hat{c}\dg\big) \;\! 
	     e^{{\cal L}_{\rm het}\;\!\tau} \Big[ \Big( \rt{\frac{\hbar\;\!\eta}{2}} \hat{c}- i \hbar \hat{f}_1 \Big) \rho + {\rm Hc} \Big] \Big\}}
	    {\Tr \Big\{ \big(\hat{c}+\hat{c}\dg\big) \;\! 
	     e^{{\cal L}_{\rm het}\;\!\tau} \Big[ \Big( -i\rt{\frac{\hbar\;\!\eta}{2}} \hat{c}- i \hbar \hat{f}_2 \Big) \rho + {\rm Hc} \Big] \Big\}}
	    {\Tr \Big\{ -i\big(\hat{c}-\hat{c}\dg\big) \;\! 
	     e^{{\cal L}_{\rm het}\;\!\tau} \Big[ \Big( -i\rt{\frac{\hbar\;\!\eta}{2}} \hat{c}- i \hbar \hat{f}_2 \Big) \rho + {\rm Hc} \Big] \Big\}}  \nn \\
    &\hspace{0.5cm} + \hbar^2 \, \ID{2} \, \delta(\tau)  \;.
\end{align}
As with the homodyne case we can derive a heterodyne quantum Langevin equation assuming $\vop{s}$ to be a scalar-operator by taking the Markovian limit from \eqref{NonzeroDelayQLE}. The result is
\begin{align}
	\hbar \, d\hat{s} = {}& \big( \, i \;\! [\hat{H}_{1},\hat{s}\;\!] + {\cal J}[\hat{c}\dg] \;\! \hat{s} - \half \;\! \{ \hat{c}\dg \hat{c}, \hat{s} \} 
	                        + \hbar \, {\cal D}[\hat{f}] \;\! \hat{s} \;\! \big) dt  \nn \\
                        & + \big[\;\!\hat{c}\dg d\hat{B}_{\rm in} - d\hat{B}\dg_{\rm in} \hat{c}, \hat{s}\;\!\big] 
	                        - i \,\rt{\frac{\hbar\;\!\eta}{2}} \big[\hat{s},\hat{F}\dg\;\!\big] \big( \hat{c} \, dt + d\hat{B}_{\rm in} \big) 
	                        + i \,\rt{\frac{\hbar\;\!\eta}{2}} \big( \hat{c}\dg \, dt + d\hat{B}\dg_{\rm in} \big) \big[ \hat{F},\hat{s}\;\!\big] \nn \\
                        & - i \big[\hat{s},\hat{F}+\rt{\bar{\eta}} F\dg\;\!\big] d\hat{U}_{\rm in1}  
                          - i \big[\hat{s}, -i( \hat{F} + \rt{\bar{\eta}} F\dg )\;\!\big] d\hat{U}_{\rm in2}  
                          + i d\hat{U}\dg_{\rm in1} \big[ \rt{\bar{\eta}}\hat{F} + \hat{F}\dg, \hat{s}\;\!\big]  
                          + i d\hat{U}\dg_{\rm in2} \big[ -i( \rt{\bar{\eta}}\hat{F} - \hat{F}\dg ), \hat{s}\;\!\big]  \;.
\end{align}
\end{widetext}

\section{Discussion}
\label{Discussion}

We have constructed a theory of Markovian quantum feedback control for nonlinear systems with an arbitrary number of decay channels, inputs, outputs, and mediated by arbitrary diffusive measurements. We have derived the time evolution of the system state both with and without conditioning, for a vacuum bath input. When the evolution is unconditioned one may find an equivalent formulation in terms of quantum Langevin equations and we have derived these equations too. We also derived the two-time correlation function for the measured current including feedback.

We have performed our derivations using the \hei\ picture, where the entire feedback loop is described by unitary evolution. Most notably we established relation \eqref{SchHeiPicRelnWithCond}, which can be viewed as the analogue of \eqref{SchHeiPicReln} but for conditional evolution. This is what allowed us to derive the stochastic master equation from the \hei\-picture quantum Langevin equations.

It is interesting to note that the two-time correlation function of the measured current is an expression about measurements at two separated times. It therefore lends itself as a different way of deriving the stochastic master equation. In this approach one would calculate the correlation function in the \sch\ picture by making the ansatz \eqref{AnsatzSME} and equating the end result to \eqref{CorrFuncnWithFeedback3}. Solving for $\vog{\alpha}$ should result in \eqref{AlphaWithFeedback}. If one was only interested in the stochastic master equation then this second method is however much less direct than the first approach, as the calculation of the correlation function in the \hei\ picture is a very lengthy process. Alternatively, one could derive a stochastic master equation first and then use it to derive the autocorrelation of the current on which the state is conditioned in the \sch\ picture. However, our approach to obtaining the autocorrelation of the current and the stochastic master equation, has not been to derive one result from the other, but rather each result independently.

The interpretation of the \hei\ picture approach was recognized in Ref.~\cite{Wis94} and also discussed in detail in Ref.~\cite{WM10}. In essence this is a no-measurement (or more precisely no-collapse) model where the observer is never aware of the measurement record from the monitoring. Consequently we have refrained from using terms such as ``unravellings'' or ``conditional'' unless in explicit reference to results in the \sch\ picture.

Finally we note that it would be possible to generalize the results of this paper even further by allowing the bath to be non-vacuum. In such a theory we would have to allow $\dBin$ to have a non-zero mean and correlated more generally as opposed just \eqref{QuantumItoRule}.

\begin{acknowledgments}

This research was conducted by the \emph{Australian Research Council Centre of Excellence for Quantum Computation and Communication Technology} (project number CE110001029).

\end{acknowledgments}

\appendix

\section{CONNECTION TO CONTROL SYSTEMS ENGINEERING}
\label{EngineeringDefintion}

The standard engineering approach to feedback control is to start with a stochastic differential equation of a vector $\bf x$. A model of the system in the time domain is known as a state-space model and the vector ${\bf x}$ a state. The state contains variables such that if these variables are known at time $t$ then all other system variables at time $t$ may be calculated from it \cite{HJS08}.

The state-space model can be translated to quantum dynamics most easily via the \hei\ picture. In the \hei\ picture we define a Hermitian vop $\hx$, from which an arbitrary system operator $\hat{s}$ can be defined. We will be considering continuous Markov processes in which case,
\begin{align}
\label{NonlinearSysConfigQLE}
	d\vop{x} = \alpha(\vop{x},\vop{u},t) \, dt + \beta(\vop{x},t) \, d\vop{v}_{\rm p} \;,
\end{align}
where $\vop{u}$ is the input (potentially arising from feedback), and $d\vop{v}_{\rm p}$ is a quantum Wiener increment defined by 
\begin{align}
	\langle d\vop{v}_{\rm p}(t) \rangle = {\bf 0} \;,
\end{align}
and the \ito\ rules
\begin{align}
	d\vop{v}_{\rm p}(t) \, d\vop{v}\tp_{\rm p}(t') = {}& 0 \quad  \forall \; t \ne t' \;, \\	
	d\vop{v}_{\rm p}(t) \, d\vop{v}\tp_{\rm p}(t) = {}& {\rm I} \, dt \;.
\end{align}
Note that $\alpha$ is a vop-valued function while $\beta$ maps to a matrix (which in general may mop-valued).

We will assume the system to be monitored via $R$ channels and that the measurement noise to be diffusive. Let us denote the measurement results by $\vop{y}_1\,$, which can be written in the general form
\begin{align}
\label{NonlinearMmtOutputOperator}
	\vop{y}_1 \, dt = g(\vop{x},\vop{u},t) \, dt + d\vop{v}_{\rm m} \;.
\end{align}
The noise term $d\vop{v}_{\rm m}$ is another Wiener increment and is what defines the measurement to be diffusive. It is often assumed that $d\vop{v}_{\rm p}$ is uncorrelated with $d\vop{v}_{\rm m}$. One could of course drop this assumption and allow the two noises to be correlated if necessary \cite{WM10}. It is conventional (and we will follow this convention) to call $\vop{y}_1$ the output.

Equation \eqref{NonlinearSysConfigQLE} is generated by a Hamiltonian which one often writes in the general form
\begin{align}
	\hat{H} = \Hext + \Hm + \Hfb  \;.
\end{align}
Here $\Hm$ is still defined by \eqref{MmtHamiltonian} but the feedback Hamiltonian is kept general, of the form, 
\begin{align}
\label{GeneralFeedbackHamiltonian}
	\Hfb = \vop{f}\tp \vop{u} \;,
\end{align}
where $\vop{u}$ and $\vop{f}$ are Hermitian and $\lceil \vop{f},\vop{u} \rfloor = 0$ to ensure the Hermiticity of $\Hfb$. The input is then used to influence some system observable $\vop{f}$. Note that when $\vop{u}$ is a feedback input it will be a functional of the output $\vop{y}_1$, which is a bath vop so the condition $\lceil \vop{f},\vop{u} \rfloor = 0$ will be guaranteed.

When the input is chosen to be linear in the output
\begin{align}
	\vop{u}(t) = L \, \ym(t-\tau) \;,
\end{align}
where $\tau$ is the feedback delay, the feedback is said to be proportional, or, Markovian (provided $\tau \to 0^+$). Note that to obtain Markovian system evolution the matrix $L$ needs to be independent of time. We will absorb $L$ into the definition of $\vop{f}$ and just define Markovian feedback by 
\begin{align}
	\vop{u}(t) = \ym(t-\tau) \;,
\end{align}
and $\vop{f}$ a $R \times 1$ vector-operator. This will keep our calculation simpler, without the need to write out $L$ explicitly. Taking the input and output to be of the same dimension is no less general than if they were of different dimensions as we can always pad zeros in $\vop{u}$ (or $\ym$, since $\vop{u}$ is just the time-delayed version of $\ym$) if there is no feedback in some of the input channels. It is also not sensible to allow $\vop{u}$ (and therefore $\ym$) to have more than $R$ components since then the inner-product $\vop{f}\tp\vop{u}$ is undefined.

\section{Derivation of \erf{dsDefinition}}
\label{LangevinEqnExplained}

It would be most natural to derive the inifinitesimal evolution given by \eqref{dsDefinition} with the full unitary operator 
\begin{align}
	\Umfb(t+dt,t) = \exp\!\big[ \!-\! i \;\! \big( \Hext + \Hm + \Hfb \big) \, dt / \, \hbar \;\!\big]  \;,
\end{align}
where $\Hm$ and $\Hfb$ are given by \eqref{MmtHamiltonian} and \eqref{requiredfbHamtau} respectively. Expanding this to order $dt$,
\begin{align}
	\Umfb(t+dt,t) = {}& \hat{1} - i \, \frac{dt}{\hbar} \, \big( \Hext + \Hm + \Hfb \big) \nn \\
	                    & - \frac{1}{2 \;\! \hbar^2} \, \big( \Hm \, dt + \Hfb \, dt \big)^2  \;.
\end{align}
The important step here is to note that cross terms between $\Hm$ and $\Hfb$ do not contribute for a nonzero feedback delay $\tau$:  
\begin{align}
	(\Hm \, dt) \;\! (\Hfb \, dt) = {}& \big( \dBin\dg \, \hc - \hc\dg \dBin \big) \big[ \hbar \, \vop{f}\tp \ym(t-\tau) \, dt \big] \\
                           	    = {}& \hbar \, \hc\tp \big[ \dBin\ddg \, \ym\tp(t-\tau) \;\! dt \big] \vop{f}  \nn \\ 
                      	            & - \hbar \, \hc\dg \big[ \dBin \, \ym\tp(t-\tau) \;\! dt \big] \vop{f} \;.
\end{align}
Recall that $\ym(t-\tau)\;\!dt$ is defined in terms of $\dBmt(t-\tau)$ and $\dBmt\ddg(t-\tau)$, which for $\tau > 0$,
\begin{align}
	\mbrac{\dBmt(t-\tau)}{\dBin(t)} = \mbrac{\dBmt(t-\tau)}{\dBin\ddg(t)} = 0 \;,
\end{align}
and similarly with $\dBmt$ replaced by $\dBmt\ddg$. Therefore the products $\dBin\ddg\,\ym\tp(t-\tau)\;\!dt$ and $\dBin\,\ym\tp(t-\tau)$ can always be written as normally ordered functions in the input fields which average to zero for a vacuum bath. Similarly, $(\Hfb \, dt) \;\! (\Hm \, dt)$ is also negligible. Letting $\hbar \equiv 1$ for simplicity, we thus obtain
\begin{align}
	& \hspace{-0.5cm} \Umfb\dg(t+dt,t) \, \vop{s} \, \Umfb(t+dt,t) \nn \\
	= {}& \vop{s} - i \, \vop{s} \, \big( \Hext + \Hm + \Hfb \big) dt \nn \\
	    & - \half \, \vop{s} \, \big( \Hm \, dt + \Hfb \, dt \big)^2 
        + i \, \big( \Hext + \Hm + \Hfb \big) dt \: \vop{s} \nn \\
      & + \big( \Hm \, dt + \Hfb \, dt \big) \, \vop{s} \, \big( \Hm \,dt + \Hfb \,dt \big) \nn \\
      & - \half \, \big( \Hm \, dt + \Hfb \, dt \big)^2 \, \vop{s}  \;.
\end{align}
Expanding and collecting terms proportional to $\Hext + \Hm$ as one group and terms proportional to $\Hfb$ as another group we get
\begin{align}
	& \hspace{-0.5cm} \Umfb\dg(t+dt,t) \, \vop{s} \, \Umfb(t+dt,t) \nn \\
%	= {}& \vop{s} + \Big[ - i \, \vop{s} \, \big( \Hext + \Hm \big) dt  + i \big( \Hext + \Hm \big) dt \: \vop{s} \nn \\
%%	
%	    & - \half \, \vop{s} \, \big( \Hm \, dt \big)^2 - \half \, \big( \Hm \, dt \big)^2 \, \vop{s} 
%	      + \big( \Hm \, dt \big) \;\! \vop{s} \;\! \big( \Hm \, dt \big) \Big]  \nn \\
%%	      
%      & + \Big[ - i \, \vop{s} \, \big( \Hfb \, dt \big)  + i \big( \Hfb \, dt \big) \, \vop{s}  \nn \\
%%        
%      & - \half \, \vop{s} \, \big( \Hfb \, dt \big)^2 - \half \, \big( \Hfb \, dt \big)^2 \, \vop{s}  
%        + \big( \Hfb \, dt \big) \;\! \vop{s} \;\! \big( \Hfb \, dt \big) \Big]  \nn \\[0.2cm]
%
%
	= {}& \vop{s} + \Big[ e^{i(\Hext + \Hm )dt} \, \vop{s} \, e^{-i(\Hext + \Hm)dt} - \vop{s} \;\!\Big] \nn \\
	    & + \Big[ e^{i \Hfb dt} \, \vop{s} \, e^{-i \Hfb dt} - \vop{s} \;\!\Big]  \;.
\end{align}
We have noted that adding $\Hext$ to $\Hm$ on the exponent of the exponential only has an effect to order $dt$. Subtracting $\vop{s}$ from each side this is simply
\begin{align}
\label{MmtPlusFeedbackQLE}
	d\vop{s} = [d\vop{s}]_{\rm m} + [d\vop{s}]_{\rm fb}  \;.
\end{align}

It should be apparent from the above that the validity of \eqref{MmtPlusFeedbackQLE} relies on the procedure of first allowing $\tau \ne 0$ and then letting $\tau \to 0^+$ in the end.

\section{DERIVATION OF \erf{FBMELindbladForm}}
\label{LindbladFBMasterEq}

We wish to derive \eqref{FBMELindbladForm} from \eqref{DiffMediatedFBME}. For convenience we restate \eqref{DiffMediatedFBME} here       
\begin{align}
\label{genMarkovFBMEAppendix}
	{\cal L}{\rho} = {}& - i \big[\hat{H}_1,\rho \big] + {\cal D}[\hat{\bf c}]\rho + \hbar \;\! {\cal D}[\hat{\bf f}]\rho  \nn \\ 
                       & - i \big\lceil \;\! \vop{f}, {\sf M}\dg \hc \;\! \rho + \rho \;\! {\sf M}\tp \hc\ddg \big\rfloor  \;.
\end{align}
Consider first the two terms ${\cal D}[\hc]\rho$ and $\big\lceil \vop{f}, {\sf M}\tp \hc \rho + \rho \;\! {\sf M}\dg \hc\ddg \big\rfloor$. Expanding the sop-bracket,
\begin{align}
\label{LfbMinusD[f]1}
	& \hspace{-0.5cm} {\cal D}[\hc] \rho - i \big\lceil \vop{f}, {\sf M}\dg \hc \rho + \rho \;\! {\sf M}\tp \hc\ddg \big\rfloor \nn \\
	=  {}& \hc\tp \! \rho \;\! \hc\ddg - \half \;\! \rho \;\! \hc\dg \hc - \half \;\! \hc\dg \hc \;\! \rho\nn \\
	     & + i \;\! \hc\!\tp \rho \;\! {\sf M}^* \;\! \vop{f} - i \;\! \vop{f}\tp {\sf M}\tp \rho \, \hc\ddg 
	       + i \rho \, \hc\dg {\sf M} \, \vop{f} - i \;\! \vop{f}\tp {\sf M}\dg \hc \, \rho  \;. 
\end{align}
We can regroup terms as follows
\begin{align}
%\label{D[c]andScalarBracket1}
	& \hspace{-0.5cm} {\cal D}[\hc] \rho - i \big\lceil \vop{f}, {\sf M}\dg \hc \rho + \rho \;\! {\sf M}\tp \hc\ddg \big\rfloor  \nn \\
%		
%	= {}& \Big( \half \;\! \hc\tp \! \rho \;\! \hc\ddg + \smallfrac{i}{2} \;\! \hc\tp \rho \;\! {\sf M}^* \;\! \vop{f} \Big) 
%	      + \smallfrac{i}{2} \;\! \hc\tp \rho \;\! {\sf M}^* \;\! \vop{f}  \nn \\
%%	
%	    & + \Big( \half \;\! \hc\tp \! \rho \;\! \hc\ddg - \smallfrac{i}{2} \;\! \vop{f}\tp {\sf M}\tp \rho \;\! \hc\ddg \Big) 
%	      - \smallfrac{i}{2} \;\! \vop{f}\tp {\sf M}\tp \rho \;\! \hc\ddg  \nn \\
%%	      
%      & - \Big( \half \;\! \rho \;\! \hc\dg \hc - \smallfrac{i}{2} \;\! \rho \;\! \hc\dg {\sf M} \;\! \vop{f} \Big) 
%        + \smallfrac{i}{2} \;\! \rho \;\! \hc\dg {\sf M} \;\! \vop{f}  \nn \\
%%      
%	    & - \Big( \half \;\! \hc\dg \hc \;\! \rho + \smallfrac{i}{2} \;\! \vop{f}\tp {\sf M}\dg \hc \;\! \rho \Big) 
%	      - \smallfrac{i}{2} \;\! \vop{f}\tp {\sf M}\dg \hc \;\! \rho  \nn \\[0.2cm] 
%		
%
%
\label{D[c]andScalarBracket2}
	= {}& \half \;\! \hc\tp \! \rho \Big( \hc\ddg + i \;\! {\sf M}^* \;\! \vop{f} \Big) 
	      + \half \Big( \hc\tp - i \;\! \vop{f}\tp {\sf M}\tp \Big) \rho \;\! \hc\ddg  \nn \\
      & - \half \;\! \rho \;\! \hc\dg \Big( \hc - i \;\! {\sf M} \;\! \vop{f} \Big)  
        - \half \Big( \hc\dg + i \;\! \vop{f}\tp {\sf M}\dg \Big) \hc \;\! \rho  \nn \\
	    & + \smallfrac{i}{2} \;\! \hc\tp \rho \;\! {\sf M}^* \;\! \vop{f} - \smallfrac{i}{2} \;\! \vop{f}\tp {\sf M}\tp \rho \;\! \hc\ddg  
        + \smallfrac{i}{2} \;\! \rho \;\! \hc\dg {\sf M} \;\! \vop{f} - \smallfrac{i}{2} \;\! \vop{f}\tp {\sf M}\dg \hc \;\! \rho  \;. 
\end{align}
%
%Now consider the commutator  
%%
%\begin{align}
%	\big[ \vop{f}\tp {\sf M}\dg \hc + \hc\dg {\sf M} \, \vop{f}, \rho \big]
%%
%	= {}& \vop{f}\tp {\sf M}\dg \hc \;\! \rho + \hc\dg {\sf M} \, \vop{f} \;\! \rho  \nn \\
%%	    
%	    & - \rho \, \vop{f}\tp {\sf M}\dg \hc - \rho \, \hc\dg {\sf M} \, \vop{f} \;, 
%\end{align}
%%
Guided by the terms with parentheses in \eqref{D[c]andScalarBracket2} we add and subtract ${\cal D}[{\sf M}\;\!\vop{f}\;\!] \rho$ to the last line in \eqref{D[c]andScalarBracket2}. Using the identity
\begin{align}
\label{RewritingCommutator}
	\smallfrac{i}{2} \, \rho \;\! \hc\dg {\sf M} \;\! \vop{f} - \smallfrac{i}{2} \, \vop{f}\tp {\sf M}\dg \;\! \hc \;\! \rho 
	= {}& \smallfrac{i}{2} \, \hc\dg {\sf M} \;\! \vop{f} \rho - \smallfrac{i}{2} \, \rho \;\! \vop{f}\tp {\sf M}\dg \hc  \nn \\
	    & - \smallfrac{i}{2} \, \big[ \vop{f}\tp {\sf M}\dg \hc + \hc\dg {\sf M} \, \vop{f}, \rho \big]  \;,
\end{align}
the last line of \eqref{D[c]andScalarBracket2} can be written as
\begin{align}
	& \hspace{-0.5cm} \smallfrac{i}{2} \;\! \hc\tp \rho \;\! {\sf M}^* \;\! \vop{f} - \smallfrac{i}{2} \;\! \vop{f}\tp {\sf M}\tp \rho \;\! \hc\ddg  
                    + \smallfrac{i}{2} \;\! \rho \;\! \hc\dg {\sf M} \;\! \vop{f} - \smallfrac{i}{2} \;\! \vop{f}\tp {\sf M}\dg \hc \;\! \rho  \nn \\
	= {}& - \smallfrac{i}{2} \big[ \vop{f}\tp {\sf M}\dg \hc + \hc\dg {\sf M} \, \vop{f}, \rho \big]  
	      + {\cal D}[{\sf M}\;\!\vop{f}\;\!] \rho - {\cal D}[{\sf M}\;\!\vop{f}\;\!] \rho  \nn \\
	    & + \smallfrac{i}{2} \, \hc\tp \, \rho \, {\sf M}^* \;\! \vop{f} - \smallfrac{i}{2} \, \vop{f}\tp {\sf M}\tp \rho \, \hc\ddg 
        + \smallfrac{i}{2} \, \hc\dg {\sf M} \;\! \vop{f} \rho - \smallfrac{i}{2} \, \rho \;\! \vop{f}\tp {\sf M}\dg \hc  \nn \\[0.2cm]
%
%
% 
%	= {}& - \smallfrac{i}{2} \big[ \vop{f}\tp {\sf M}\dg \hc + \hc\dg {\sf M} \, \vop{f}, \rho \big] - {\cal D}[{\sf M}\;\!\vop{f}\;\!] \rho \nn \\
%%	
%	    & + \Big( \smallfrac{i}{2} \, \hc\tp \rho \, {\sf M}^* \;\! \vop{f} + \half \, \vop{f}\tp {\sf M}\tp \rho \;\! {\sf M}^* \;\! \vop{f} \Big) \nn \\
%	    & - \Big( \smallfrac{i}{2} \, \vop{f}\tp {\sf M}\tp \rho \, \hc\ddg - \half \, \vop{f}\tp {\sf M}\tp \rho \;\! {\sf M}^* \;\! \vop{f} \Big) \nn \\
%	    & + \Big( \smallfrac{i}{2} \, \hc\dg {\sf M} \;\! \vop{f} \rho - \half \, \vop{f}\tp {\sf M}\dg {\sf M} \;\! \vop{f} \rho \Big) \nn \\
%	    & - \Big( \smallfrac{i}{2} \, \rho \;\! \vop{f}\tp {\sf M}\dg \hc + \half \, \rho \;\! \vop{f}\tp {\sf M}\dg {\sf M} \;\! \vop{f} \Big) 
%	      \nn \\[0.2cm]
%	    
%
%	   
	= {}& - \smallfrac{i}{2} \big[ \vop{f}\tp {\sf M}\dg \hc + \hc\dg {\sf M} \, \vop{f}, \rho \big] - {\cal D}[{\sf M}\;\!\vop{f}\;\!] \rho \nn \\
	    & + \smallfrac{i}{2} \Big( \hc\tp - i \;\! \vop{f}\tp {\sf M}\tp \Big) \rho \;\! {\sf M}^* \;\! \vop{f} 
	      - \smallfrac{i}{2} \, \vop{f}\tp {\sf M}\tp \rho\Big( \hc\ddg + i \;\! {\sf M}^* \;\! \vop{f} \Big) \nn \\
	    & + \smallfrac{i}{2} \Big( \hc\dg + i \;\! \vop{f}\tp {\sf M}\dg \Big) {\sf M} \;\! \vop{f} \rho 
	      - \smallfrac{i}{2} \, \rho \;\! \vop{f}\tp {\sf M}\dg \Big( \hc - i \;\! {\sf M} \;\! \vop{f} \Big) \;.   	      
\end{align}
Substituting this back into \eqref{D[c]andScalarBracket2} and collecting like terms we get
\begin{align}
	& \hspace{-0.5cm} {\cal D}[\hc] \rho - i \big\lceil \vop{f}, {\sf M}\dg \hc \rho + \rho \;\! {\sf M}\tp \hc\ddg \big\rfloor  \nn \\
%
%	= {}& - \smallfrac{i}{2} \big[ \vop{f}\tp {\sf M}\dg \hc + \hc\dg {\sf M} \, \vop{f}, \rho \big] - {\cal D}[{\sf M}\;\!\vop{f}\;\!] \rho \nn \\
%%	
%	    & + \Big( \hc\tp - i \;\! \vop{f}\tp {\sf M}\tp \Big) \rho \Big( \hc\ddg + i \;\! {\sf M}^* \;\! \vop{f} \Big) \nn \\
%%	      
%	    & + \smallfrac{1}{2} \, \rho \Big( \hc\dg + i \;\! \vop{f}\tp {\sf M}\dg \Big) \Big( \hc - i \;\! {\sf M} \;\! \vop{f} \Big)  \nn \\
%%	    
%	    & - \smallfrac{1}{2} \Big( \hc\dg + i \;\! \vop{f}\tp {\sf M}\dg \Big) \Big( \hc - i \;\! {\sf M} \;\! \vop{f} \Big) \rho  \\[0.2cm]
%	    
%
%
	= {}& - \smallfrac{i}{2} \big[ \vop{f}\tp {\sf M}\dg \hc + \hc\dg {\sf M} \, \vop{f}, \rho \big] 
	      + {\cal D}[\hc-i{\sf M} \;\!\vop{f}\;\!] \rho  \nn \\
      & - {\cal D}[{\sf M} \;\!\vop{f}\;\!] \rho  \;.
\end{align}
Substituting this back into ${\cal L}\;\!\rho$ we arrive at 
\begin{align}
	{\cal L} \;\! \rho = {}& - i \big[ \Hext + \half ( \;\! \vop{f}\tp {\sf M}\dg \hc + \hc\dg {\sf M} \, \vop{f} \;\! ), \rho \big]  \nn \\
	                       & + {\cal D}[\hc-i{\sf M}\;\!\vop{f}\;\!] \rho 
	                         + \hbar \;\! {\cal D}[\vop{f}] \rho - {\cal D}[{\sf M}\;\!\vop{f}\;\!] \rho \;.
\end{align}
The final two terms can be written as   
\begin{align}
\label{ProblemTermsAppendix}
	\hbar \;\! {\cal D}[\;\!\vop{f}\;\!] \rho - {\cal D}[\;\!{\sf M}\;\!\vop{f}\;\!] \rho 
	= {}& \hbar \, \vop{f}\tp \;\! \rho \;\! \vop{f} - \half \, \vop{f}\tp {\sf M}\tp \rho \;\! {\sf M}^* \;\! \vop{f}  \nn \\
	    & - \smallfrac{\hbar}{2} \, \vop{f}\tp \vop{f} \rho + \half \, \vop{f}\tp {\sf M}\dg {\sf M} \;\! \vop{f} \rho \nn \\
	    & - \smallfrac{\hbar}{2} \, \rho \;\! \vop{f}\tp \vop{f} + \half \, \rho \;\! \vop{f}\tp {\sf M}\dg {\sf M} \;\! \vop{f} \nn \\[0.2cm]
%	    
%
%
%	= {}& \vop{f}\tp \rho \;\! \big( \hbar \, \ID{R} - {\sf M}\dg {\sf M} \big)^* \;\! \vop{f}  \nn \\
%	    & - \half \, \vop{f}\tp \big( \hbar \, \ID{R} - {\sf M}\dg {\sf M} \big) \, \vop{f} \;\! \rho  \nn \\
%	    & - \half \, \rho \;\! \vop{f}\tp \big( \hbar \, \ID{R} - {\sf M}\dg {\sf M} \big) \, \vop{f}  \nn \\[0.2cm]
%
%
%
	= {}& \vop{f}\tp \rho {\sf Z}^* \;\! \vop{f} 
	      - \half \, \vop{f}\tp {\sf Z} \, \vop{f} \;\! \rho - \half \, \rho \, \vop{f}\tp {\sf Z} \;\! \vop{f} \,.
\end{align}
%
%
%
%The matrix inside the parentheses in \eqref{ProblemTermsAppendix} 
%%
%\begin{align}
%	{\sf Z} \equiv \hbar \, I - {\sf M}\dg {\sf M} \;,
%\end{align}
%%
%will be positive semidefinite if the eigenvalues of ${\sf M}\dg {\sf M}$ are bounded above by $\hbar$. Let the set of nonzero eigenvalues of a matrix $X$ be denoted by $\lambda\big(X\big)$. We note that
%%
%\begin{align}
%	\lambda\big( {\sf M}\dg {\sf M} \big) = \lambda\big( {\sf M} {\sf M}\dg \big) = \hbar \, \lambda\big( {\sf H} \big) \le \hbar  \;,
%\end{align}
%%
%Recall that ${\sf H}={\rm diag}(\bm{\eta})$ where $\eta_{k} \in [0,1]$ for every $k$. The first equality follows by considering the singular value decomposition of ${\sf M}$. The second equality follows from ${\sf M}{\sf M}\dg={\sf H}$, which one can show from \eqref{UnravellingMatrix} and \eqref{DefnOfM}. We can thus write ${\sf Z}$ in terms of its matrix square root ${\sf B}$, defined by ${\sf Z}={\sf B}\dg {\sf B}$, 
%
%
%
Recall that ${\sf Z}=\hbar\,\ID{R}-\MdgM\,$, which was defined under \eqref{DefnOfZ}. Since ${\sf Z} \ge 0\;\!$, there exists a ${\sf B}$ such that ${\sf Z} = {\sf B}\dg {\sf B}$. Therefore we are free to write
\begin{align}
	&  \hspace{-0.5cm}\hbar \;\! {\cal D}[\;\!\vop{f}\;\!] \rho - {\cal D}[\;\!{\sf M}\;\!\vop{f}\;\!] \rho  \nn \\
	= {}& \vop{f}\tp {\sf B}\tp \rho \;\! {\sf B}^* \;\! \vop{f}  
	      - \half \, \vop{f}\tp \big( {\sf B}\dg {\sf B} \big) \, \vop{f} \rho  
	      - \half \, \rho \;\! \vop{f}\tp \big( {\sf B}\dg {\sf B} \big) \, \vop{f}  \nn \\
%	    
%	= {}& \big( {\sf B} \;\! \vop{f} \big)\!\;\!\tp \rho \big( {\sf B} \;\! \vop{f} \big)\ddg 
%	      - \half \big( {\sf B} \;\! \vop{f} \big)\;\!\!\dg \big( {\sf B} \;\! \vop{f} \big) \rho  
%        - \half \rho \big( {\sf B} \;\! \vop{f} \big)\;\!\!\dg \big( {\sf B} \;\! \vop{f} \big)  \nn \\
%        
	= {}& {\cal D}[{\sf B}\;\!\vop{f}] \rho  \;.
\end{align}
The final master equation in Lindblad form is therefore (leaving ${\sf B}$ as a general matrix square root)
\begin{align}
 \hbar \;\! \dot{\rho} = {}& - i \big[ \Hext + \half ( \;\! \vop{f}\tp {\sf M}\dg \hc + \hc\dg {\sf M} \, \vop{f} \;\! ), \rho \big]  \nn \\
	                         & + {\cal D}[\hc-i{\sf M}\;\!\vop{f}\;\!] \rho + {\cal D}[{\sf B}\;\!\vop{f}\;\!] \rho  \;.
\end{align}

\section{DERIVATION OF \erf{CorrFuncnWithFeedback2}}
\label{DerivanOfCorrFuncnWithFeedback}

For clarity we will label each term in \eqref{CorrFuncnWithFeedback1}:
\begin{widetext}
\begin{align}
\label{AppendixCorrFuncnWithFeedback1}
	\hbar^2 \ban{\ymfb(t) \, \ymfb\tp(t+\tau)} 
	= {}& \underset{\text{term A}}{\ban{{\sf M}\dg \;\! \bmfb(t) \, \bmfb\tp(t+\tau) \, {\sf M}^*}}  
	      + \underset{\text{term B}}{\ban{{\sf M}\dg \;\! \bmfb(t) \, \bmfb\dg(t+\tau) \, {\sf M}}}
	      + \underset{\text{term E}}{\ban{{\sf M}\dg \;\! \bmfb(t) \, \hbar \;\! \zetamfb\tp(t+\tau)}}  \nn \\
	    & + \underset{\text{term C}}{\ban{{\sf M}\tp \;\! \bmfb\ddg(t) \, \bmfb\tp(t+\tau) \, {\sf M}^*}}
	      + \underset{\text{term D}}{\ban{{\sf M}\tp \;\! \bmfb\ddg(t) \, \bmfb\dg(t+\tau) \, {\sf M}}} 
	      + \underset{\text{term F}}{\ban{{\sf M}\tp \;\! \bmfb\ddg(t) \, \hbar \;\! \zetamfb\tp(t+\tau)}}  \nn \\
      & + \underset{\text{term G}}{\ban{\hbar \, \zetamfb(t) \, \bmfb\tp(t+\tau) \, {\sf M}^*}}  
        + \underset{\text{term H}}{\ban{\hbar \, \zetamfb(t) \, \bmfb\dg(t+\tau) \, {\sf M}}}
        + \underset{\text{term I}}{\ban{\hbar \, \zetamfb(t) \, \hbar \, \zetamfb\tp(t+\tau)}} \,.
\end{align}
For convenience we use ``cw'' to abbreviate ``cancels with''. Normal and time ordering of each term leads to
\\
Term A:
\begin{align}
	{\ban{{\sf M}\dg \;\! \bmfb(t) \, \bmfb\tp(t+\tau) \, {\sf M}^*}}
	= {}& \underset{\rm A1}{\ban{\;\!{\sf M}\dg \hc(t+\tau) \,\hc\tp\!(t) \,{\sf M}^*}\tp}
	      - \underset{\text{A2 (cw G2)}}{\ban{i\;\!{\sf M}\dg \hc(t+\tau) \,\vop{f}\tp\!(t) (\MdgM)^*}\tp}  \nn \\
	    & - \underset{\text{A3 (cw E1)}}{\ban{i\;\!\MdgM \,\vop{f}(t+\tau) \,\hc\tp\!(t) {\sf M}^*}\tp} 
	      - \underset{\text{A4 (cw G4)}}{\ban{\MdgM \, \vop{f}(t+\tau) \, \vop{f}\tp\!(t) (\MdgM)^*}\tp}  \;.
\end{align}
Term B:
\begin{align}
	\ban{{\sf M}\dg \;\! \bmfb(t) \, \bmfb\dg(t+\tau) \, {\sf M}}
	= {}& \underset{\rm B1}{\ban{{\sf M}\tp \hc\ddg(t+\tau) \, \hc\tp\!(t) \, {\sf M}^*}\tp} 
	      - \underset{\text{B2 (cw H2)}}{\ban{\;\!i\, {\sf M}\tp \hc\ddg(t+\tau) \, \vop{f}\tp\!(t) \,(\MdgM)^*}\tp}  
	      + \underset{\text{B5 (cw I1)}}{\hbar^2 \, \MdgM \, \delta(\tau)}  \nn \\
      & + \underset{\text{B3 (cw E3)}}{\ban{\;\!i\, (\MdgM)^* \, \vop{f}(t+\tau) \,\hc\tp(t) \, {\sf M}^*}\tp} 
        + \underset{\text{B4 (cw E4)}}{\ban{(\MdgM)^* \, \vop{f}(t+\tau) \, \vop{f}(t) \, (\MdgM)^*}\tp}  \;.
\end{align}
Term C:
\begin{align}
	\ban{{\sf M}\tp \;\! \bmfb\ddg(t) \, \bmfb\tp(t+\tau) \, {\sf M}^*}
	= {}& \underset{\rm C1}{\ban{{\sf M}\tp \hc\ddg(t) \, \hc\tp\!(t+\tau) \, {\sf M}^*}} 
	      - \underset{\text{C2 (cw F1)}}{\ban{\;\!i\, {\sf M}\tp \hc\ddg(t) \, \vop{f}\tp\!(t+\tau) \, (\MdgM)^*}}  \nn \\
	    & + \underset{\text{C3 (cw G6)}}{\ban{\;\!i\,(\MdgM)^* \, \vop{f}(t) \, \hc\tp\!(t+\tau) \, {\sf M}^*}}
	      + \underset{\text{C4 (cw F2)}}{\ban{(\MdgM)^* \, \vop{f}(t) \, \vop{f}\tp\!(t+\tau) \, (\MdgM)^*}}  \;.
\end{align}
Term D:
\begin{align}
	\ban{{\sf M}\tp \;\! \bmfb\ddg(t) \, \bmfb\dg(t+\tau) \, {\sf M}}
	= {}& \underset{\rm D1}{\ban{{\sf M}\tp \hc\ddg(t) \, \hc\dg(t+\tau) \, {\sf M}}}
	     + \underset{\text{D2 (cw F3)}}{\ban{\;\!i\, {\sf M}\tp \hc\ddg(t) \, \vop{f}\tp(t+\tau) \, \MdgM}}  \nn \\
     & + \underset{\text{D3 (cw H6)}}{\ban{\;\!i\, (\MdgM)^* \, \vop{f}(t) \, \hc\dg(t+\tau) \, {\sf M}}}
       - \underset{\text{D4 (cw F4)}}{\ban{(\MdgM)^* \, \vop{f}(t) \, \vop{f}\tp(t+\tau) \, \MdgM}}  \;.
\end{align}
Term E:
\begin{align}
	\ban{{\sf M}\dg \;\! \bmfb(t) \, \hbar \;\! \zetamfb\tp(t+\tau)} 
	= {}& \underset{\text{E1 (cw A3)}}{\ban{\;\!i\,\MdgM \, \vop{f}(t+\tau) \, \hc\tp\!(t) \, {\sf M}^*}\tp}
	      + \underset{\text{E2 (cw I3)}}{\ban{\MdgM \, \vop{f}(t+\tau) \, \vop{f}\tp\!(t) \, (\MdgM)^*}\tp}  \nn \\
      & - \underset{\text{E3 (cw B3)}}{\ban{\;\!i\, (\MdgM)^* \, \vop{f}(t+\tau) \, \hc\tp(t) \, {\sf M}^*}\tp}
        - \underset{\text{E4 (cw B4)}}{\ban{(\MdgM)^* \, \vop{f}(t+\tau) \, \vop{f}\tp\!(t) \, (\MdgM)^*}\tp}  \;.
\end{align}
Term F:
\begin{align}
	\ban{{\sf M}\tp \;\! \bmfb\ddg(t) \, \hbar \;\! \zetamfb\tp(t+\tau)} 
	= {}& \underset{\text{F1 (cw C2)}}{\ban{\;\!i\,{\sf M}\tp \hc\ddg(t) \, \vop{f}\tp\!(t+\tau) \, (\MdgM)^*}}
	      - \underset{\text{F2 (cw C4)}}{\ban{(\MdgM)^* \, \vop{f}(t) \, \vop{f}\tp\!(t+\tau) \, (\MdgM)^*}}  \nn \\
      & - \underset{\text{F3 (cw D2)}}{\ban{\;\!i\, {\sf M}\tp \hc\ddg(t) \, \vop{f}\tp\!(t+\tau) \, \MdgM}}
        + \underset{\text{F4 (cw D4)}}{\ban{(\MdgM)^* \, \vop{f}(t) \, \vop{f}\tp\!(t+\tau) \, \MdgM}}  \;.
\end{align}
Term G:
\begin{align}
	\ban{\hbar \, \zetamfb(t) \, \bmfb\tp(t+\tau) \, {\sf M}^*}
	= {}& \underset{\rm G1}{\ban{-i\;\!\hbar\, {\sf M}\dg \hc(t+\tau) \, \vop{f}\tp\!(t)}\tp} 
	      + \underset{\text{G2 (cw A2)}}{\ban{\;\!i\, {\sf M}\dg \hc(t+\tau) \, \vop{f}\tp\!(t) \, (\MdgM)^*}\tp}  \nn \\
	    & - \underset{\text{G3 (cw I2)}}{\ban{\hbar \, \MdgM \, \vop{f}(t+\tau) \, \vop{f}\tp\!(t)}\tp}
	      + \underset{\text{G4 (cw A4)}}{\ban{\MdgM \, \vop{f}(t+\tau) \, \vop{f}\tp\!(t) \, (\MdgM)^*}\tp}  	\nn \\
      & + \underset{\rm G5}{\ban{\;\!i\;\!\hbar \, \vop{f}(t) \, \hc\tp\!(t+\tau) \, {\sf M}^*}} 
        - \underset{\text{G6 (cw C3)}}{\ban{\;\!i\, (\MdgM)^* \, \vop{f}(t) \, \hc\tp\!(t+\tau) \, {\sf M}^*}}  \nn \\
      & + \underset{\text{G7 (cw I6)}}{\ban{\hbar \, \vop{f}(t) \, \vop{f}\tp\!(t+\tau) \, (\MdgM)^*}}
        - \underset{\text{G8 (cw I7)}}{\ban{(\MdgM)^* \, \vop{f}(t) \, \vop{f}\tp\!(t+\tau) (\MdgM)^*}}  \;.
\end{align}
Term H:
\begin{align}
	\ban{\hbar \, \zetamfb(t) \, \bmfb\dg(t+\tau) \, {\sf M}}
	= {}& \underset{\rm H1}{\ban{-i\;\!\hbar\, {\sf M}\tp \hc\ddg(t+\tau) \, \vop{f}\tp\!(t)}\tp}
		    + \underset{\text{H2 (cw B2)}}{\ban{i\,{\sf M}\tp \hc\ddg(t+\tau) \, \vop{f}\tp\!(t+\tau) \, (\MdgM)^* }\tp}  \nn \\
      & + \underset{\text{H3 (cw I4)}}{\ban{\hbar\, (\MdgM)^* \, \vop{f}(t+\tau) \, \vop{f}\tp\!(t)}\tp}
        - \underset{\text{H4 (cw I5)}}{\ban{(\MdgM)^* \, \vop{f}(t+\tau) \, \vop{f}\tp\!(t) \, (\MdgM)^*} \tp}  \nn \\
	    & + \underset{\rm H5}{\ban{\;\!i\;\!\hbar\, \vop{f}(t) \, \hc\dg(t+\tau) {\sf M}}}
	      - \underset{\text{H6 (cw D3)}}{\ban{\;\!i\, (\MdgM)^* \, \vop{f}(t) \, \hc\dg(t+\tau) \, {\sf M}}}  \nn \\
      & - \underset{\text{H7 (cw I8)}}{\ban{\hbar\, \vop{f}(t) \, \vop{f}\tp\!(t+\tau) \, \MdgM}} 
        + \underset{\text{H8 (cw I9)}}{\ban{(\MdgM)^* \, \vop{f}(t) \, \vop{f}\tp\!(t+\tau) \, \MdgM}}  \;.
\end{align}
Term I:
\begin{align}
	\ban{\hbar \, \zetamfb(t) \, \hbar \, \zetamfb\tp(t+\tau)} 
	= {}& \hbar^2 \, \ID{R} \, \delta(\tau) - \underset{\text{I1 (cw B5)}}{\hbar \, \MdgM \, \delta(\tau)}  \nn \\
      & + \underset{\text{I2 (cw G3)}}{\ban{\hbar\, \MdgM \, \vop{f}(t+\tau) \, \vop{f}\tp\!(t)}\tp}
        - \underset{\text{I3 (cw E2)}}{\ban{\MdgM \, \vop{f}(t+\tau) \, \vop{f}\tp(t) \, (\MdgM)^*}\tp}  \nn \\
      & - \underset{\text{I4 (cw H3)}}{\ban{\hbar (\MdgM)^* \, \vop{f}(t+\tau) \, \vop{f}\tp\!(t)}\tp}
        + \underset{\text{I5 (cw H4)}}{\ban{(\MdgM)^* \vop{f}(t+\tau) \, \vop{f}\tp\!(t) \, (\MdgM)^*}\tp}  \nn \\
      & - \underset{\text{I6 (cw G7)}}{\ban{\hbar\, \vop{f}(t) \, \vop{f}\tp\!(t+\tau) \, (\MdgM)^*}}
        + \underset{\text{I7 (cw G8)}}{\ban{(\MdgM)^* \, \vop{f}(t) \, \vop{f}\tp\!(t+\tau) \, (\MdgM)^*}}  \nn \\
      & + \underset{\text{I8 (cw H7)}}{\ban{\hbar\, \vop{f}(t) \, \vop{f}\tp(t+\tau) \, \MdgM}}   
        + \underset{\text{I9 (cw H8)}}{\ban{(\MdgM)^* \, \vop{f}(t) \, \vop{f}\tp(t+\tau) \, \MdgM}}  \;.
\end{align}
\end{widetext}
The remaining terms are A1, B1, C1, D1, G1, G5, H1, H5, and the $\hbar^2\,\ID{R}\,\delta(\tau)$ in term I. Adding these and collecting like terms we arrive at \eqref{CorrFuncnWithFeedback2}.

\end{document}